\documentclass{aa}

\usepackage{graphicx}

\newcommand{\feh}{\mathrm{[Fe/H]}}
\newcommand{\teff}{T_\mathrm{eff}}
\newcommand{\thetaeff}{\theta_\mathrm{eff}}
\newcommand{\logg}{\log g}

\begin{document}

\title{IRFM $T_\mathrm{eff}$ calibrations for cluster and field giants in
the Vilnius, Geneva, $RI_\mathrm{(C)}$ and DDO photometric systems
\thanks{Based on data from the GCPD.}}

\author{I. Ram\'{\i}rez \inst{1,2} \and J. Mel\'endez \inst{1,3}}

\offprints{I. Ram\'{\i}rez \\ \email{ivan@astro.as.utexas.edu}}

\institute{Seminario Permanente de Astronom\'{\i}a y Ciencias
Espaciales, Universidad Nacional Mayor de San Marcos, \\ Ciudad
Universitaria, Facultad de Ciencias F\'{\i}sicas, Av. Venezuela
s/n, Lima 1, Per\'u \and Department of Astronomy, The University
of Texas at Austin, RLM 15.202A, TX 78712-1083 \and Department of
Astronomy, California Institute of Technology, MC 105--24,
Pasadena, CA 91125}

\date{Received - - - / Accepted - - -}

\abstract{Based on a large sample of disk and halo giant stars,
for which accurate effective temperatures derived through the
InfraRed Flux Method (IRFM) exist, a calibration of the
temperature scale in the Vilnius, Geneva, $RI_\mathrm{(C)}$ and
DDO photometric systems is performed. We provide calibration
formulae for the metallicity dependent $\teff$ vs color relations
as well as grids of intrinsic colors and compare them with other
calibrations. Photometry, atmospheric parameters and reddening
corrections for the stars of the sample have been updated with
respect to the original sources in order to reduce the dispersion
of the fits. Application of our results to Arcturus leads to an
effective temperature in excellent agreement with the value
derived from its angular diameter and integrated flux. The effects
of gravity on these $\teff$ vs color relations are also explored
by taking into account our previous results for dwarf stars.
}

\authorrunning{I. Ram\'{\i}rez and J. Mel\'endez}
\titlerunning{IRFM temperature calibrations for giants}

\maketitle

\section{Introduction}

In a previous paper (Mel\'endez \& Ram\'{\i}rez \cite{melendez03},
Paper~I) we derived $\teff:$ color $:\feh$ relations for dwarf
stars in the Vilnius, Geneva, $RI_\mathrm{(C)}$ and DDO systems
from the Alonso et~al. (\cite{alonso96a}) sample. This time, we
have performed a similar extension for giants by using the
corresponding Alonso et~al. (\cite{alonso99a}) sample. We have
chosen this sample in order to maintain the homogeneity of the
calibrations, the effective temperatures of all of the stars in
both samples (dwarfs and giants) have been obtained in a single
implementation of the InfraRed Flux Method (IRFM) by Alonso and
coworkers.

In addition to their primary importance as fundamental relations,
these empirical $\teff$ calibrations are extremely useful for
several purposes. Applications include chemical abundance studies
(e.g. Smith et~al. \cite{smith02}, Kraft \& Ivans \cite{kraft03}),
the transformation of theoretical HR diagrams into their
observational counterparts, i.e. color-magnitude planes (e.g.
Girardi et~al. \cite{girardi02}) and the test of synthetic spectra
and colors (e.g. Bell \cite{bell97}). When used along with other
studies, combined results may have implications in galactic
chemical evolution, population synthesis and cosmology.

A noteworthy feature of our work and that of Alonso et~al.
(\cite{alonso96b}, \cite{alonso99b}) is the inclusion of
Population II stars in the calibrations. They allow to extend the
ranges of applicability of the formulae, which in turn become
useful also for metal-poor stars. In order to achieve this goal, a
large number of cluster giants has been included in the sample
adopted for the present work. Even though for a given photometric
system observations are usually available for only 2 or 3
clusters, the number of stars contributed by each of them is
considerable and thus deserved to be included.

When comparing temperature scales of main sequence stars with
those corresponding to giants, small but systematic differences
arise. Theoretical calculations can easily take into account these
variations with the $\logg$ value. In empirical studies, however,
it is only after a considerable number of stars has been studied
and their properties properly averaged that gravity effects become
clear. Detailed, careful inspection of gravity effects on colors
may improve our understanding of the physics behind stellar
spectra formation.

The present work is distributed as follows: Sect.
\ref{secc:...adopted} describes the data adopted, in Sect.
\ref{secc:...cal} we discuss some properties of the general
calibration formula employed, perform the calibrations and apply
them to Arcturus. Comparison of our work with previously published
calibrations is presented in Sect. \ref{secc:...comp} while
intrinsic colors of giant stars and the effects of gravity on the
temperature scale are discussed in Sect. \ref{secc:...escala}. We
finally summarize our results in Sect.~\ref{secc:...conclusion}.

\section{Photometry and atmospheric parameters adopted} \label{secc:...adopted}

Colors adopted in this work were obtained from the General
Catalogue of Photometric Data (GCDP, Mermilliod et~al.
\cite{mermilliod97}), as in Paper I. Nevertheless, due to the lack
of $RI_\mathrm{(C)}$ photometry for giants, colors obtained by
applying transformation equations to Kron-Eggen and Washington
photometry were also adopted, as explained in Sect.
\ref{sect:ric}.

The reddening corrections $E(B-V)$ are given in the work of Alonso
et~al. (\cite{alonso99a}). For most of the stars in globular
clusters, however, these values have been updated following Kraft
\& Ivans (\cite{kraft03}, KI03). Although reddening ratios
$E(\mathrm{color})/E(B-V)$ for medium and broad band systems
generally depend on spectral type, their variations with
luminosity class for our interval of interest (F5-K5) amount to a
maximum of 0.05 and so only relevant values have been changed with
respect to the values adopted for dwarfs. Thus, for giants we have
taken $E(Z-V)/E(B-V)=0.30$ and $E(Y-Z)/E(B-V)=0.54$ (Strai\v{z}ys
\cite{straizys95}). The reddening ratio $E(B_1-B_2)/E(B-V)=0.35$
(Cramer \cite{cramer99}) was also necessary to obtain intrinsic
$t$ parameters in the Geneva system (Sect. \ref{sect:geneva}). For
the remaining colors, the adopted reddening ratios are the same as
those given in Paper I (see our Table 1 and references therein).

Several stars were discarded before performing the fits due to
their anomalous positions in $\teff$ vs color planes. It is worth
mentioning the most discrepant of them and the probable causes of
this behaviour (temperatures are those given by Alonso et~al.
\cite{alonso99a}): BD +11 2998, noted by Alonso et~al.
(\cite{alonso99a}) as a ``probable misindentification in the
program of near IR photometry'', an F8 star with very high
$\teff=7073$ K; and BS 2557, also HD 50420, a blue A9III variable
star, $\teff=4871$ K.

Regarding $\feh$ values, the catalogue of Cayrel de Strobel et~al.
(\cite{cayrel92}) has been superseded by the 2001 edition (Cayrel
de Strobel et~al. \cite{cayrel01}). In addition, we have updated
the 2001 catalogue from several abundance studies, introducing
more than 1000 entries and 357 stars not included in the original
catalogue. The most important sources for this update were:
Mishenina \& Kovtyukh (\cite{mishenina01}), Santos et~al.
(\cite{santos01}), Heiter \& Luck (\cite{heiter03}), Stephens \&
Boesgaard (\cite{stephens02}), Takada-Hiday et~al.
(\cite{takada02}) and Yong \& Lambert (\cite{yong03}). We will
call ``C03'' this actualized catalogue.

\begin{figure}
 \includegraphics[bb=19 10 567 311,width=8.7cm]{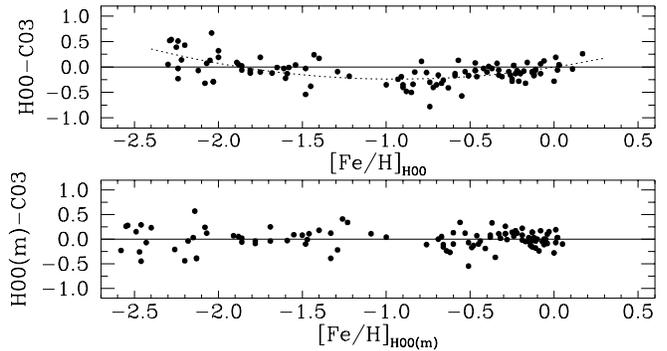}
 \caption{Upper panel: Differences between photometric $\feh$
 values from the Hilker (\cite{hilker00}, H00) calibration and C03
 spectroscopic determinations for a subsample of giant stars.
 Bottom panel: Differences between the modified photometric
 $\feh$ values [H00(m)] and C03 spectroscopic determinations.}
 \label{fig:h00vsc03}
\end{figure}

Hilker (\cite{hilker00}, H00) has published a metallicity
calibration for giants in the Str\"{o}mgren system, which, when
compared with the spectroscopic metallicities of C03 shows clear
systematic tendencies (Fig. \ref{fig:h00vsc03}). The systematic
difference between H00 and C03 fit approximately the following
empirical relation:
$\feh_\mathrm{H00}-\feh_\mathrm{C03}=0.274\feh^{2}+0.511\feh-0.002$,
which is also shown in Fig. \ref{fig:h00vsc03} as a dotted line.
By substracting this difference from the $\feh$ values obtained
from the H00 calibration we obtain a modified H00 metallicity in
better agreement with C03.

We have adopted $\feh$ values in order of reliability: firstly C03
values ($\sim70\%$ of the sample), secondly the modified Hilker
(\cite{hilker00}) calibration ($\sim5\%$) and lastly the
photometric metallicities given by Alonso et~al.
(\cite{alonso99a}).

\section{The calibrations} \label{secc:...cal}

Broadly speaking, the behaviour of the $\teff$ vs color relations
can be represented by the simple equation:
$\teff=c_1/(c_2\mathrm{X}+c_3)$, where X is the color index and
$c_1$, $c_2$ and $c_3$ constants (see for example Hauck \&
K\"{u}nzli \cite{hauck96}). It is easy to show, by aproximating
the stellar flux to that of a blackbody, that this equation has
some physical meaning. To better reproduce the observed gradients
$\Delta\teff/\Delta\mathrm{X}$ and to take into account the
effects of different chemical compositions, however, the following
general formula has proved to be more accurate:
\begin{eqnarray}
\thetaeff & = & a_0+a_1\mathrm{X}+a_2\mathrm{X}^{2}+a_3\mathrm{X}\feh \nonumber \\
& & +a_4\feh+a_5\feh^{2}\ . \label{gen-eq}
\end{eqnarray}
Here, $\thetaeff=5040/\teff$ and $a_i$ ($i=0,1,\ldots,5$) are the
constants of the fit (see for example Alonso et~al.
\cite{alonso96b}, \cite{alonso99b}; Mel\'endez \& Ram\'irez
\cite{melendez03}).

Nonlinear fits of the data to Eq. (\ref{gen-eq}) were performed
for 7 color indices and 1 photometric parameter as described in
the following subsections. Coefficients such that $3\sigma
(a_i)>a_i$ were neglected and stars departing more than
$2.5\sigma(\teff)$ from the mean fit where iteratively discarded
(the number of iterations hardly exceeded 5).

It has been argued that the $a_5$ term, which is almost always
negative, systematically produces too high temperatures for
metal-poor stars (Ryan et~al. \cite{ryan99}, Nissen et~al.
\cite{nissen02}). For a non-negligible $a_5$ term, the dependence
of $\teff$ on $\feh$ gradually increases as very low values of
$\feh$ are considered. Physically, this is not an expected
behaviour since $\teff$ should become nearly independent of $\feh$
as less metals are present in the stellar atmosphere. We have
carefully checked the residuals of every iteration in order to
reduce this kind of tendencies, specially for the
$-3.0<\feh\leq-2.5$ group, and have made $a_5=0$ when necessary.

Even after these considerations, the residuals of some fits showed
systematic tendencies, specially for the stars with $\feh>-1.5$.
This may be due to the fact that the hydrogen lines, the G band
and continuum discontinuities as the Paschen jump fall into some
of the bandpasses. In order to remove these small tendencies, we
fitted the original residuals to high order polynomials and
substracted them from the $5040/\thetaeff=f($color$,\feh)$ values
derived at first. Therefore, the temperature of a star is to be
obtained according to
\[\teff=5040/\thetaeff(\mathrm{color},\feh)-P(\mathrm{color},\feh)\ ,\]
where $P$ is the polynomial fit to the original residuals. The
final calibration formulae are thus not as practical as the
original ones but much more accurate than them. Interpolation from
the resulting $\teff$ vs color vs $\feh$ Tables
\ref{vilnius:intrinsic-colors}, \ref{geneva:intrinsic-colors} and
\ref{ric-ddo:intrinsic-colors} constitute a more practical
approach to the effective temperature of a star.

\subsection{Vilnius system}

\begin{figure*}
 \centering
 \includegraphics[bb=19 10 563 310,width=17cm]{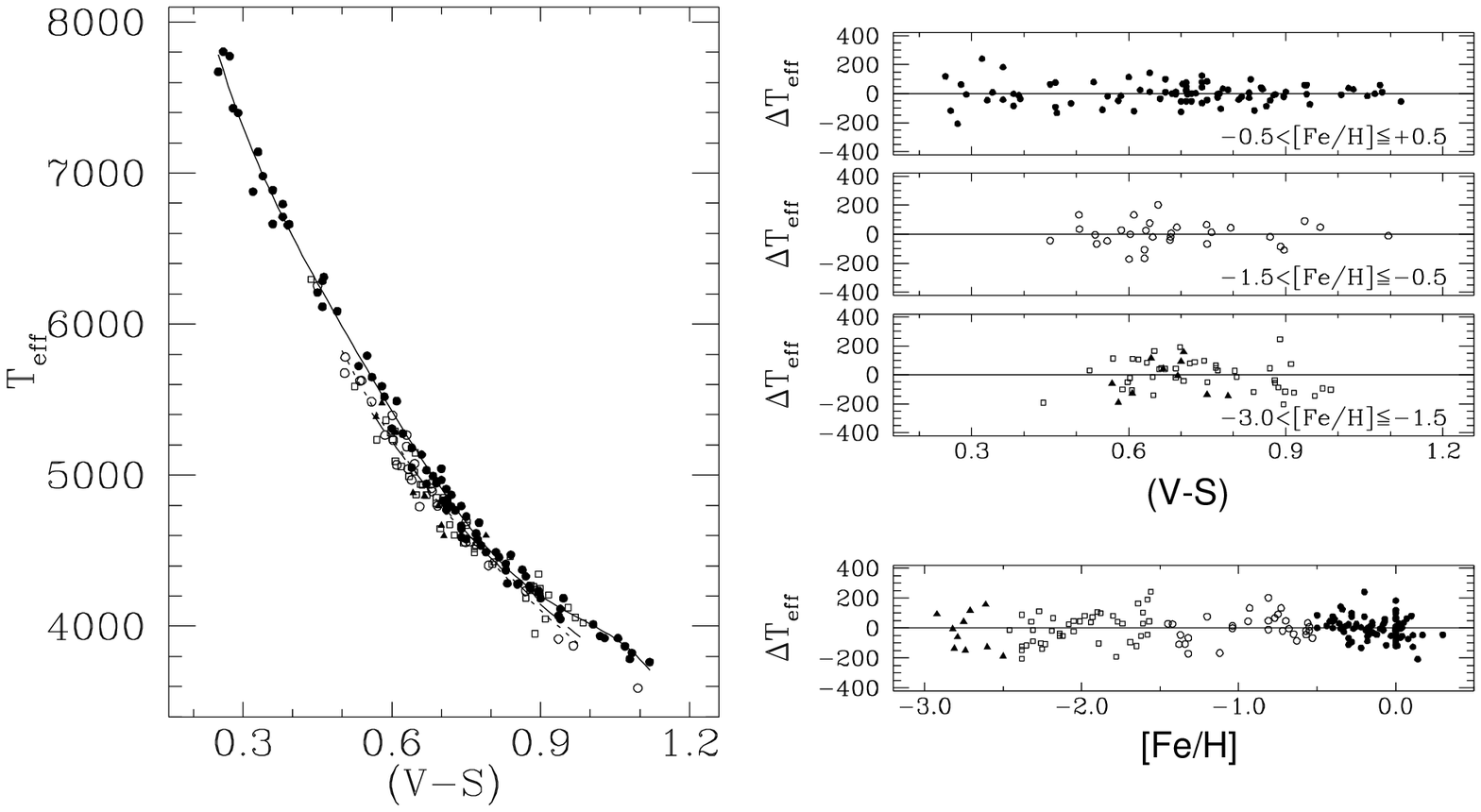}
 \caption{Left: $\teff$ vs $(V-S)$ observed for the metallicity
   ranges $-0.5<\feh\leq+0.5$ (filled circles), $-1.5<\feh\leq-0.5$
   (open circles), $-2.5<\feh\leq-1.5$ (squares) and
   $-3.0<\feh\leq-2.5$ (triangles). Curves corresponding to our
   calibration for $\feh=0$ (solid line), $\feh=-1$ (dotted line) and
   $\feh=-2$ (dashed line) are also shown. Right: residuals of the
   fit ($\Delta\teff=\teff^\mathrm{cal}-\teff^\mathrm{IRFM}$)
   as a function of $(V-S)$ (for the metallicity ranges indicated in
   the lower right section of the three upper panels) and $\feh$
   (bottom panel).}
 \label{fig:vilvs}
\end{figure*}

The filters defining the colors we have calibrated in this system
are (approximate effective wavelengths and bandwidths in nm,
according to Strai\v{z}ys \& Sviderskiene \cite{straizys72}, are
given in parenthesis): $Y$ (466, 26), $V$ (544, 26) and $S$~(655,
20).

For the $(V-S)$ color index we found:
\begin{equation}
\thetaeff=0.440+0.838(V-S)-0.011\feh\ , \label{eq:vs}
\end{equation}
with $\sigma(\teff)=86$ K (after correcting with the polynomial
fits to the original residuals given in Table \ref{table:polfits})
and $N=170$. Hereafter $\sigma(\teff)$ and $N$ will be used to
denote the standard deviation in $\teff$ and the number of stars
included in the fit, respectively.

The sample and residuals of this fit are shown in
Fig.~\ref{fig:vilvs}, from which we see that Eq. (\ref{eq:vs}) is
applicable in the following ranges:
\begin{eqnarray}
+0.25<(V-S)<+1.12 & \textrm{ for} & -0.5<\feh<+0.5, \nonumber\\
+0.50<(V-S)<+0.98 & \textrm{ for} & -1.5<\feh<-0.5, \nonumber\\
+0.56<(V-S)<+0.98 & \textrm{ for} & -2.5<\feh<-1.5, \nonumber\\
+0.57<(V-S)<+0.79 & \textrm{ for} & -3.0<\feh<-2.5. \nonumber
\end{eqnarray}

Most of the stars included in this and the following fits have
temperatures between 4000 K and 5000 K. A considerable number of
giant stars in globular clusters belong to this last group.

Around $(V-S)=0.30$ the sensitivity of this color to $\teff$ is
such that $\Delta\teff/\Delta(V-S)\simeq90$ K per 0.01 mag. This
sensitivity gradually decreases for cool stars reaching 20 K per
0.01 mag at $(V-S)=1.10$. On the other hand, the mean variation
$\Delta\teff/\Delta\feh$ is almost independent of $\feh$ and
varies only slightly with color from 20 K per 0.3 dex at
$(V-S)\simeq0.50$ to 10 K per 0.3 dex at $(V-S)\simeq0.90$.

\begin{figure*}
 \centering
 \includegraphics[bb=19 10 563 310,width=17cm]{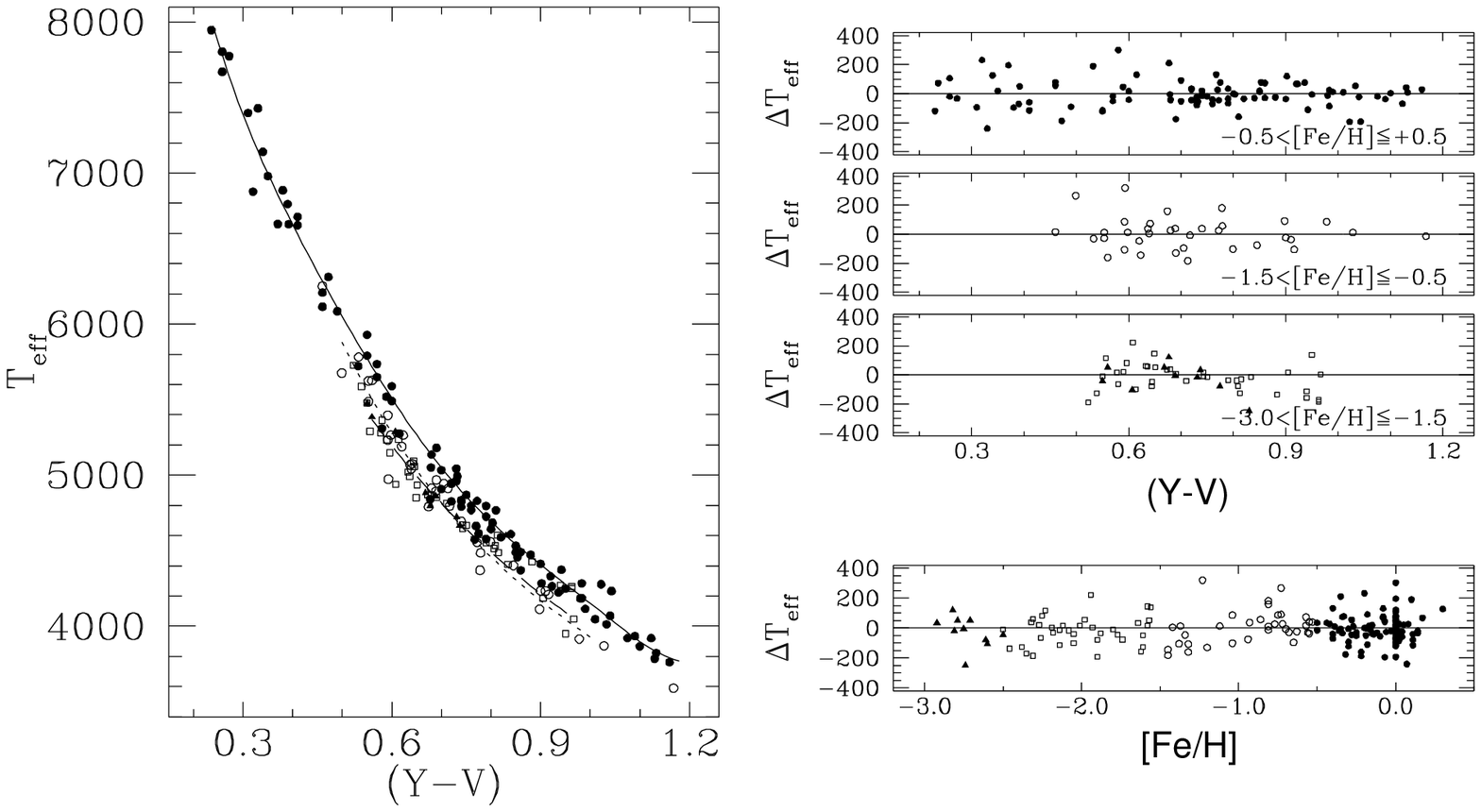}
 \caption{As in Fig. \ref{fig:vilvs} for $\teff$ vs $(Y-V)$.}
 \label{fig:vilyv}
\end{figure*}

For the $(Y-V)$ color index, data satisfy the following formula:
\begin{eqnarray}
\thetaeff & = & 0.416+0.900(Y-V)-0.099(Y-V)^{2} \nonumber\\
& & -0.059\feh-0.017\feh^{2}
 \label{eq:yv}
\end{eqnarray}
with $\sigma(\teff)=99$ K and $N=172$. This is, again, after the
residuals correction (Table \ref{table:polfits}).

The ranges of applicability of Eq. (\ref{eq:yv}) can be inferred
from Fig. \ref{fig:vilyv}, where we show the sample and residuals
of the fit:
\begin{eqnarray}
+0.24<(Y-V)<+1.18 & \textrm{ for} & -0.5<\feh<+0.5, \nonumber\\
+0.50<(Y-V)<+1.00 & \textrm{ for} & -1.5<\feh<-0.5, \nonumber\\
+0.56<(Y-V)<+0.96 & \textrm{ for} & -2.5<\feh<-1.5. \nonumber \\
+0.56<(Y-V)<+0.75 & \textrm{ for} & -3.0<\feh<-2.5. \nonumber
\end{eqnarray}

The sensitivity of $(Y-V)$ to the effective temperature is very
similar to that of the $(V-S)$ color index. There is, however, a
greater influence of $\feh$ on this relation. The gradient
$\Delta\teff/\Delta\feh$ depends on $\feh$, reaching zero as
$\feh\longrightarrow-2$. A typical value of
$\Delta\teff/\Delta\feh$ for solar metallicity stars is 90 K per
0.3 dex.

\subsection{Geneva system} \label{sect:geneva}

Our calibrations for this system span a considerable range of
temperatures, going approximately from 3600 K to 8200~K for solar
metallicity stars. All the fits for the Geneva colors need to be
corrected using Table \ref{table:polfits}.

\begin{figure*}
 \centering
 \includegraphics[bb=19 10 563 310,width=17cm]{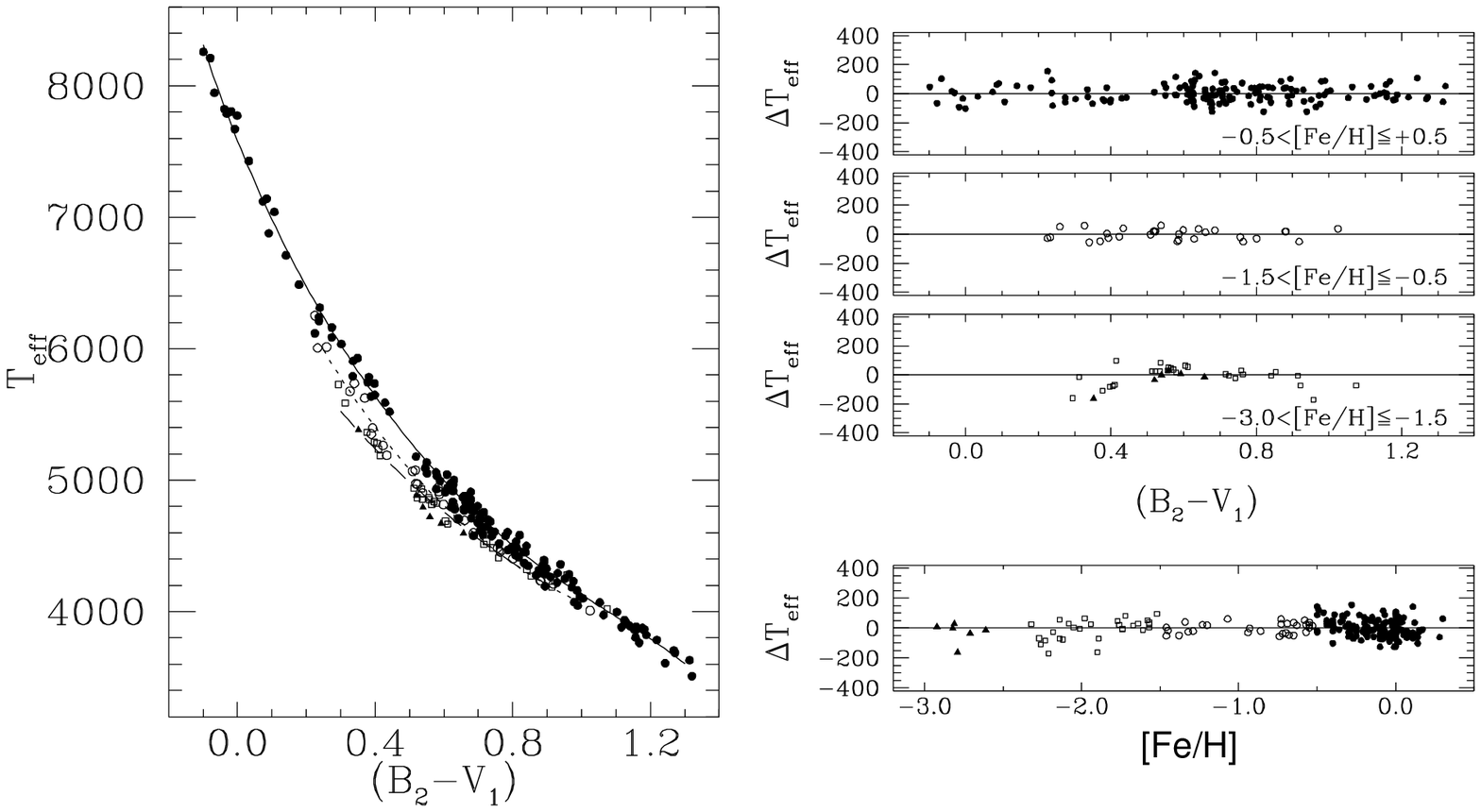}
 \caption{As in Fig. \ref{fig:vilvs} for $\teff$ vs $(B_2-V_1)$.}
 \label{fig:genb2v1}
\end{figure*}

For the $\teff:(B_2-V_1):\feh$ relation we found the following
fit:
\begin{eqnarray}
\thetaeff & = & 0.662+0.604(B_2-V_1)-0.040(B_2-V_1)^{2} \nonumber \\
& & +0.039(B_2-V_1)\feh-0.048\feh \label{eq:genb2v1}
\end{eqnarray}
with $\sigma(\teff)=57$ K and $N=230$.

Figure \ref{fig:genb2v1} shows the sample and residuals of this
fit, which is valid in the following ranges:
\begin{eqnarray}
-0.10<(B_2-V_1)<+1.30 & \textrm{ for} & -0.5<\feh<+0.5, \nonumber\\
+0.25<(B_2-V_1)<+1.00 & \textrm{ for} & -1.5<\feh<-0.5, \nonumber\\
+0.30<(B_2-V_1)<+0.90 & \textrm{ for} & -2.5<\feh<-1.5, \nonumber\\
+0.52<(B_2-V_1)<+0.66 & \textrm{ for} & -3.0<\feh<-2.5. \nonumber
\end{eqnarray}

\begin{figure*}
 \centering
 \includegraphics[bb=19 10 563 310,width=17cm]{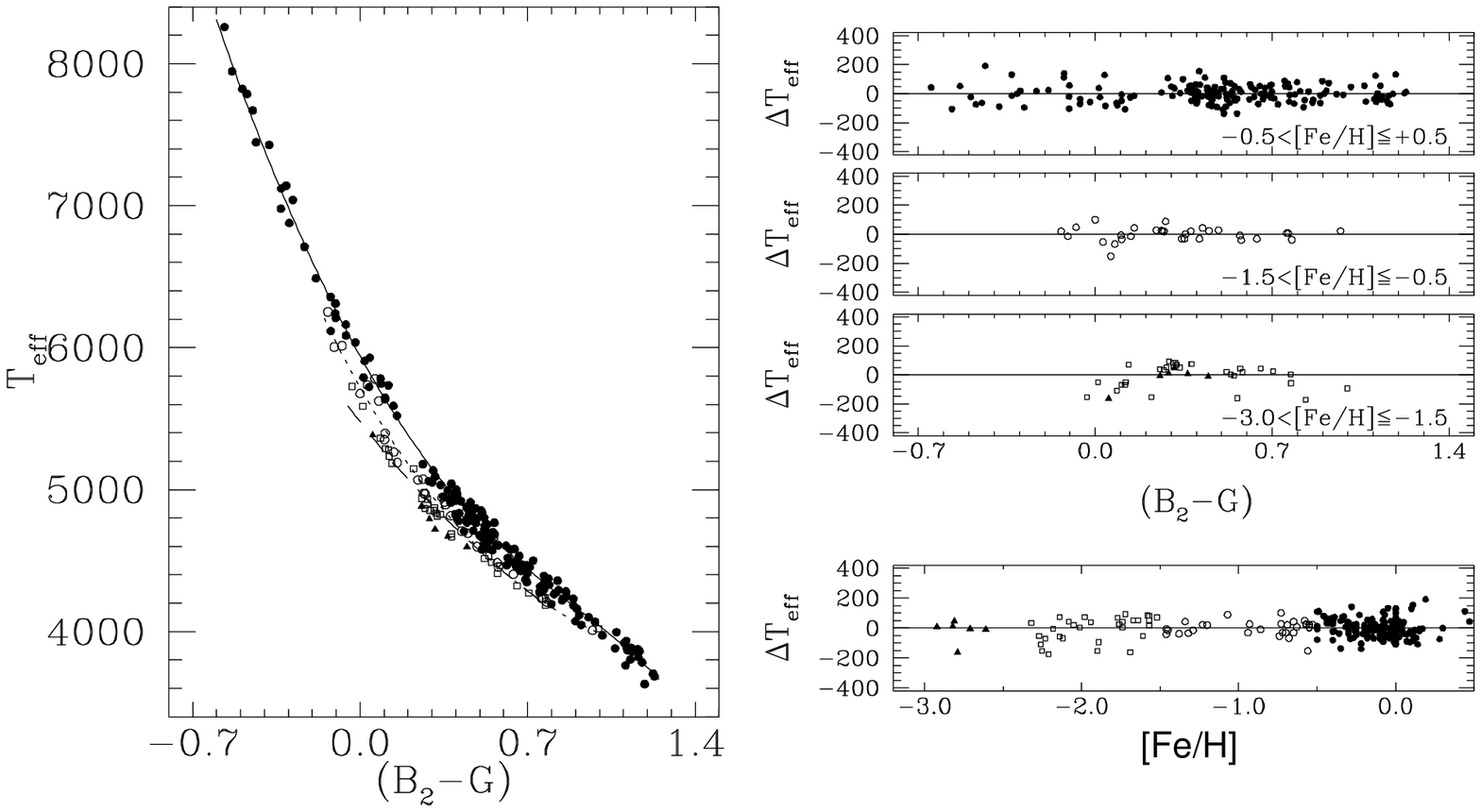}
 \caption{As in Fig. \ref{fig:vilvs} for $\teff$ vs $(B_2-G)$.}
 \label{fig:genb2g}
\end{figure*}

Likewise, for the $(B_2-G)$ color we obtained:
\begin{eqnarray}
\thetaeff & = & 0.852+0.408(B_2-G) \nonumber \\
& & +0.021(B_2-G)\feh-0.034\feh \label{eq:genb2g}
\end{eqnarray}
with $\sigma(\teff)=62$ K and $N=235$.

Eq. (\ref{eq:genb2g}) is valid in the following ranges (see Fig.
\ref{fig:genb2g}):
\begin{eqnarray}
-0.60<(B_2-G)<+1.25 & \textrm{ for} & -0.5<\feh<+0.5, \nonumber\\
-0.15<(B_2-G)<+1.00 & \textrm{ for} & -1.5<\feh<-0.5, \nonumber\\
-0.05<(B_2-G)<+0.80 & \textrm{ for} & -2.5<\feh<-1.5, \nonumber\\
+0.26<(B_2-G)<+0.48 & \textrm{ for} & -3.0<\feh<-2.5. \nonumber
\end{eqnarray}

Gradients $\Delta\teff/\Delta(B_2-V_1)$ and
$\Delta\teff/\Delta(B_2-G)$ corresponding to Eqs.
(\ref{eq:genb2v1}) and (\ref{eq:genb2g}) range from about 70~K per
0.01 mag for the hottest stars to 15 K per 0.01 mag for the cool
end. The color index $(B_2-V_1)$ is only slightly better than
$(B_2-G)$ as a $\teff$ indicator provided that the star
metallicity is known. The influence of this last parameter on
$\teff$, according to our calibration formulae (\ref{eq:genb2v1})
and (\ref{eq:genb2g}), is high for stars with $\teff\sim5800$ K
($\Delta\teff/\Delta\feh\simeq70$~K per 0.3 dex at
$(B_2-V_1)\simeq0.30$ and $(B_2-G)\simeq-0.10$) and gradually
disappears as cooler stars are considered .

It is worth mentioning that it was hard to fit the points for
stars having $-1.2<\feh<-0.8$ and 4800~K$<\teff<5000$ K, basically
due to the low number of stars included around $\feh\sim-1.0$,
which correspond to the transition region between halo and disk
(see Figs. \ref{fig:genb2v1}, \ref{fig:genb2g}). In fact, there
are few stars with abundance determinations in the interval
$-1.2<\feh<-0.8$ (see e.g. Fig.~5c in Mel\'endez \& Barbuy
\cite{melendez02}).

A photometric parameter nearly independent of $\feh$ in the Geneva
system  is the $t$ parameter, defined as
$t\equiv(B_2-G)-0.39(B_1-B_2)$ (Strai\v{z}ys \cite[p.
372]{straizys95}). Compared to the dwarf calibration, which
disperses when cool dwarfs are included, the $t$ parameter for
giant stars covers a greater range of temperatures due to its
higher sensitivity to $\teff$ for cool giants.

\begin{figure*}
 \centering
 \includegraphics[bb=19 10 563 310,width=17cm]{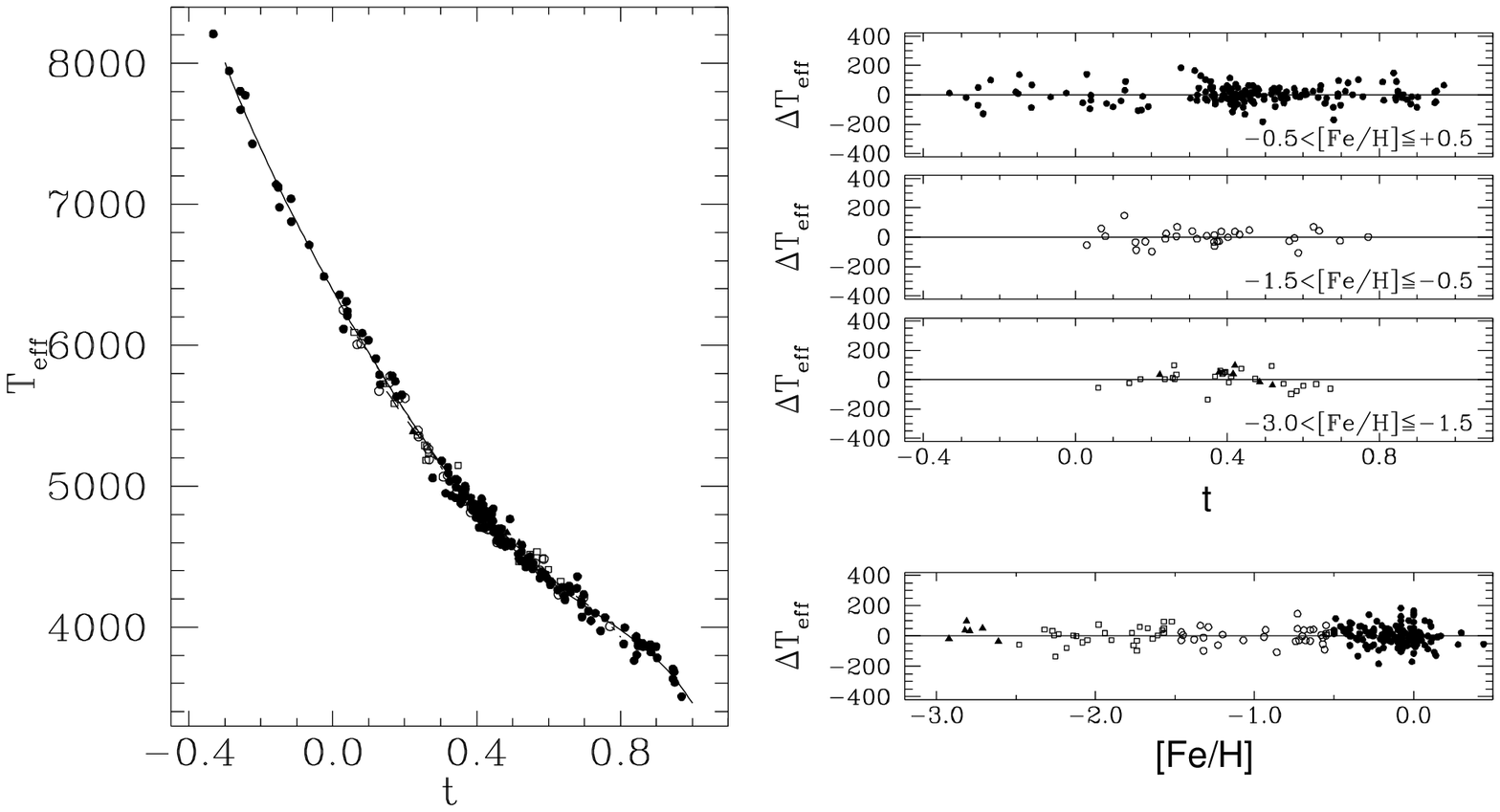}
 \caption{As in Fig. \ref{fig:vilvs} for $\teff$ vs $t$.}
 \label{fig:gent}
\end{figure*}

Data for the $t$ parameter satisfy the following fit:
\begin{equation}
\thetaeff=0.800+0.600t+0.005t\feh \label{eq:gent}
\end{equation}
with $\sigma(\teff)=59$ K and $N=228$.

Eq. (\ref{eq:gent}) is applicable in the following ranges:
\begin{eqnarray}
-0.30<t<+1.00 & \textrm{ for} & -0.5<\feh<+0.5, \nonumber\\
+0.05<t<+0.80 & \textrm{ for} & -1.5<\feh<-0.5, \nonumber\\
+0.15<t<+0.65 & \textrm{ for} & -2.5<\feh<-1.5, \nonumber\\
+0.22<t<+0.52 & \textrm{ for} & -3.0<\feh<-2.5. \nonumber
\end{eqnarray}

Even though the mean variations $\Delta\teff/\Delta(B_2-V_1)$ and
$\Delta\teff/\Delta(B_2-G)$ are similar to $\Delta\teff/\Delta t$,
the independence of $\feh$ for the $\teff$ vs $t$ relation makes
the $t$ parameter a better $\teff$ indicator. The mean variation
$\Delta\teff/\Delta\feh$ corresponding to Eq. (\ref{eq:gent}) is
always lower than 10 K per 0.3 dex.

\subsection{$RI_\mathrm{(C)}$ system} \label{sect:ric}

Only 15 \% of the stars in the sample have $RI_\mathrm{(C)}$
photometry available in the GCPD. Approximately, for another 15~\%
we compiled Kron-Eggen and Washington photometry. By applying the
transformation formulae of Bessell (\cite{bessell79},
\cite{bessell01}), we derived $(R-I)_\mathrm{(C)}$ and
$(V-I)_\mathrm{(C)}$ values from these other colors.

\begin{figure*}
 \centering
 \includegraphics[bb=19 10 563 310,width=17cm]{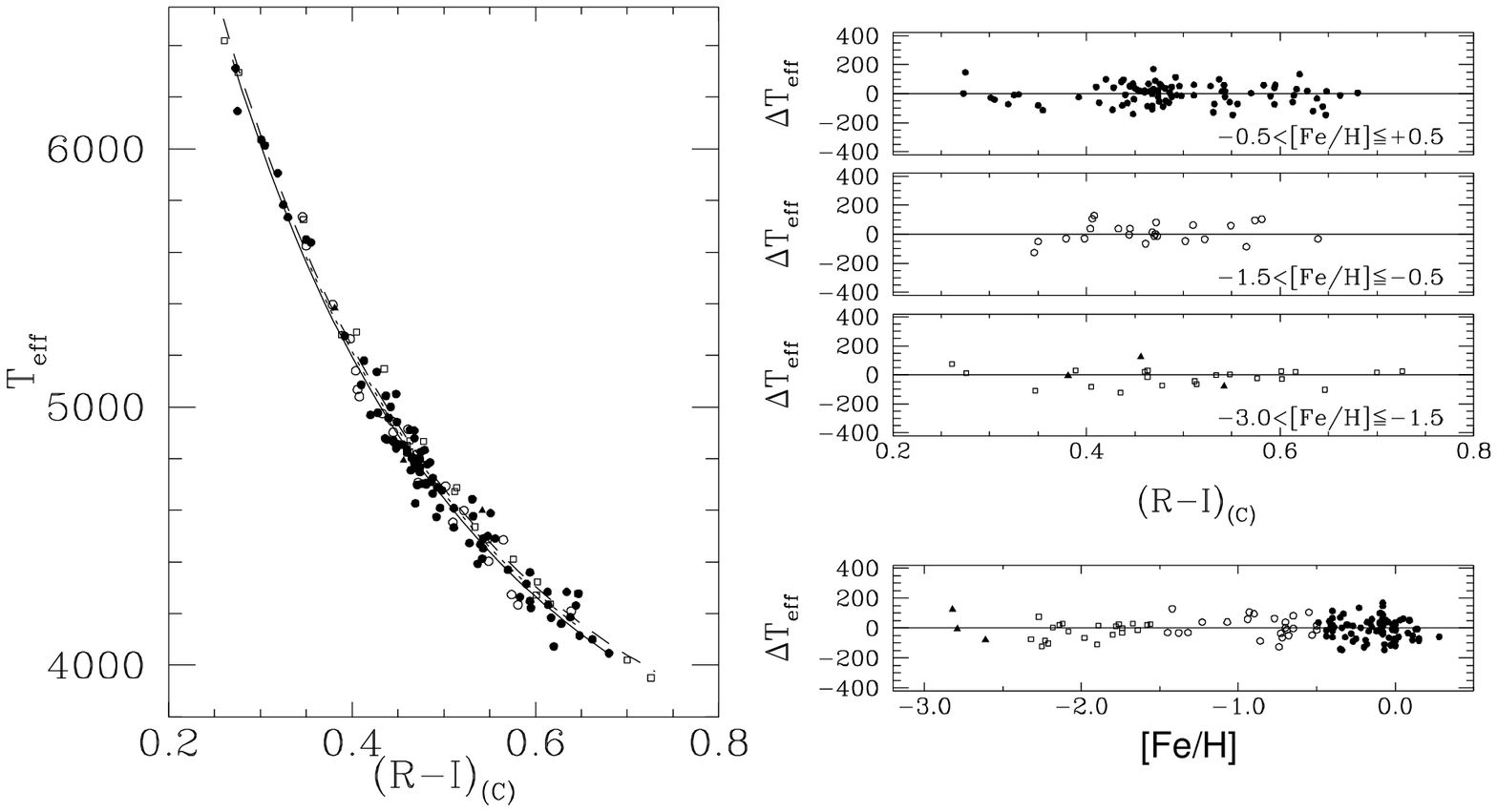}
 \caption{As in Fig. \ref{fig:vilvs} for
          $\teff$ vs $(R-I)_\mathrm{(C)}$.}
 \label{fig:ric}
\end{figure*}

Data for this system satisfy the following fit:
\begin{eqnarray}
\thetaeff & = & 0.332+1.955(R-I)_\mathrm{(C)} \nonumber \\
& & -0.898(R-I)_\mathrm{(C)}^{2}+0.009\feh \label{eq:ric}
\end{eqnarray}
with $\sigma(\teff)=67$ K and $N=137$. No appreciable tendencies
were found in the residuals, so there is no need for a residual
fit here.

The sample and residuals of this fit are shown in Fig.
\ref{fig:ric}, which shows that Eq. (\ref{eq:ric}) is applicable
only in the following ranges:
\begin{eqnarray}
+0.27<(R-I)_\mathrm{(C)}<+0.68 & \textrm{ for} & -0.5<\feh<+0.5, \nonumber\\
+0.35<(R-I)_\mathrm{(C)}<+0.65 & \textrm{ for} & -1.5<\feh<-0.5, \nonumber\\
+0.26<(R-I)_\mathrm{(C)}<+0.73 & \textrm{ for} & -2.5<\feh<-1.5, \nonumber\\
+0.38<(R-I)_\mathrm{(C)}<+0.54 & \textrm{ for} & -3.0<\feh<-2.5.
\nonumber
\end{eqnarray}

The gradient $\Delta\teff/\Delta(R-I)_\mathrm{(C)}$, which
independently of $\feh$ amounts from 100 K per 0.01 mag at
$(R-I)_\mathrm{(C)}\simeq0.30$ to 20 K per 0.01 mag at
$(R-I)_\mathrm{(C)}\simeq0.70$ and the low values of the mean
variations $\Delta\teff/\Delta\feh$ ($<20$ K per 0.3 dex)
corresponding to Eq. (\ref{eq:ric}) make this color an excellent
temperature indicator for giants.

\begin{figure*}
 \centering
 \includegraphics[bb=19 10 563 310,width=17cm]{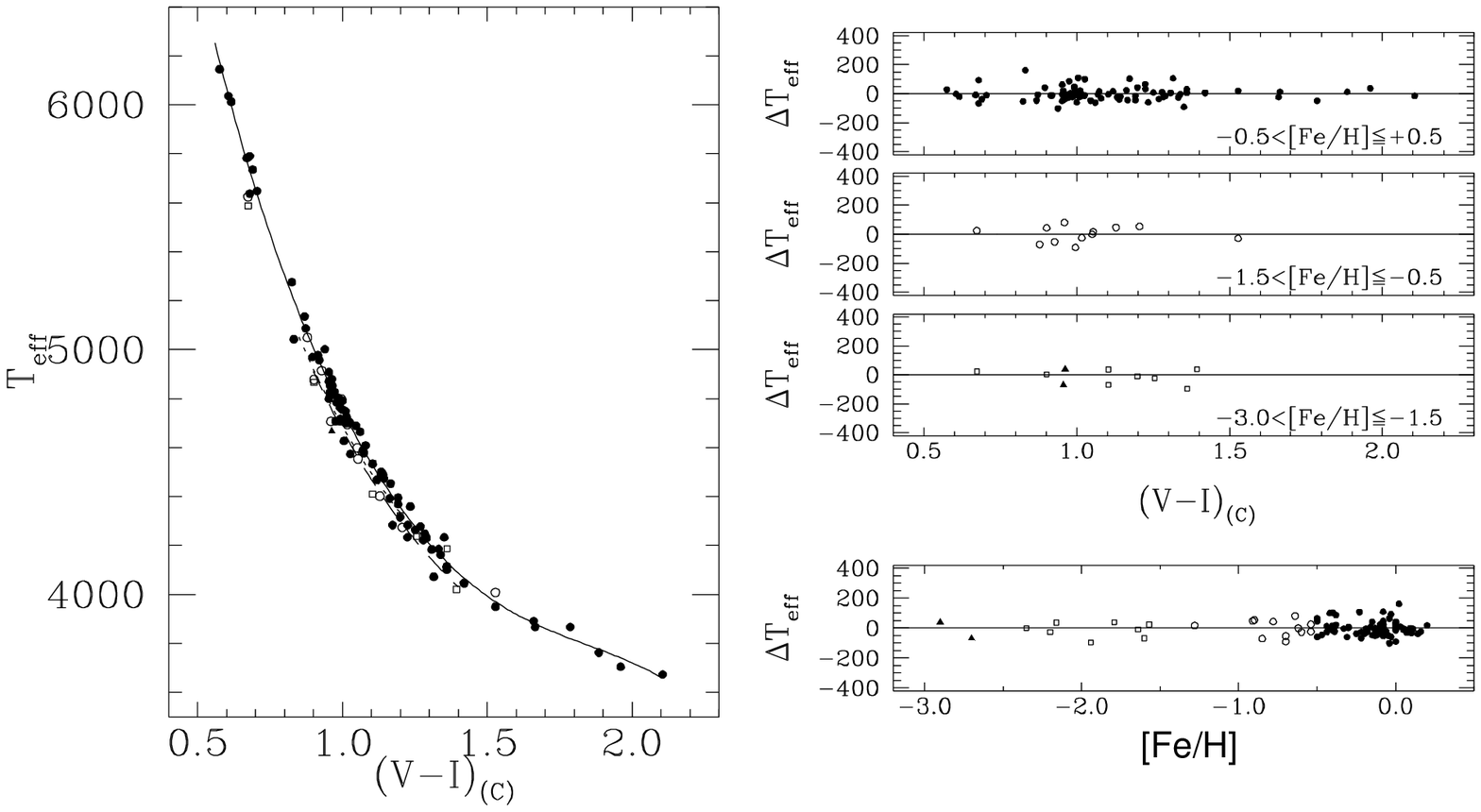}
 \caption{As in Fig. \ref{fig:vilvs} for
          $\teff$ vs $(V-I)_\mathrm{(C)}$.}
 \label{fig:ricvi}
\end{figure*}

Even better is the calibration for $(V-I)_\mathrm{(C)}$ (Fig.
\ref{fig:ricvi}):
\begin{eqnarray}
\thetaeff & = & 0.374+0.886(V-I)_\mathrm{(C)} \nonumber \\
& & -0.197(V-I)_\mathrm{(C)}^{2}-0.008(V-I)_\mathrm{(C)}\feh
\label{eq:ricvi}
\end{eqnarray}
which has $\sigma(\teff)=46$ K and $N=111$. A residual correction
is required here though (Table \ref{table:polfits}).

Eq. (\ref{eq:ricvi}) is applicable in the following ranges:
\begin{eqnarray}
+0.56<(V-I)_\mathrm{(C)}<+2.10 & \textrm{ for} & -0.5<\feh<+0.5, \nonumber\\
+0.85<(V-I)_\mathrm{(C)}<+1.20 & \textrm{ for} & -1.5<\feh<-0.5, \nonumber\\
+0.90<(V-I)_\mathrm{(C)}<+1.40 & \textrm{ for} & -2.5<\feh<-1.5.
\nonumber
\end{eqnarray}

There is a steep $\Delta\teff/\Delta(V-I)_\mathrm{(C)}$ gradient
from $(V-I)_\mathrm{(C)}=0.56$ to 1.30 for $\feh=0$, which, later
on decreases considerably to less than 10 K per 0.01 mag. The mean
variation $\Delta\teff/\Delta\feh$ is constant over the common
ranges and amounts to approximately 10 K per 0.3 dex.

\subsection{DDO system}

There is a considerable number of stars with DDO photometry
available in the GCPD though only in the range 3800 K$<\teff<$5800
K. Although DDO colors are severely affected by the star
metallicity, the calibration formula obtained for the $C(42-48)$
color satisfactorily reproduces the effect and, surprisingly, its
standard deviation is very close to the lowest to be found in this
work.

\begin{figure*}
 \centering
 \includegraphics[bb=19 10 563 310,width=17cm]{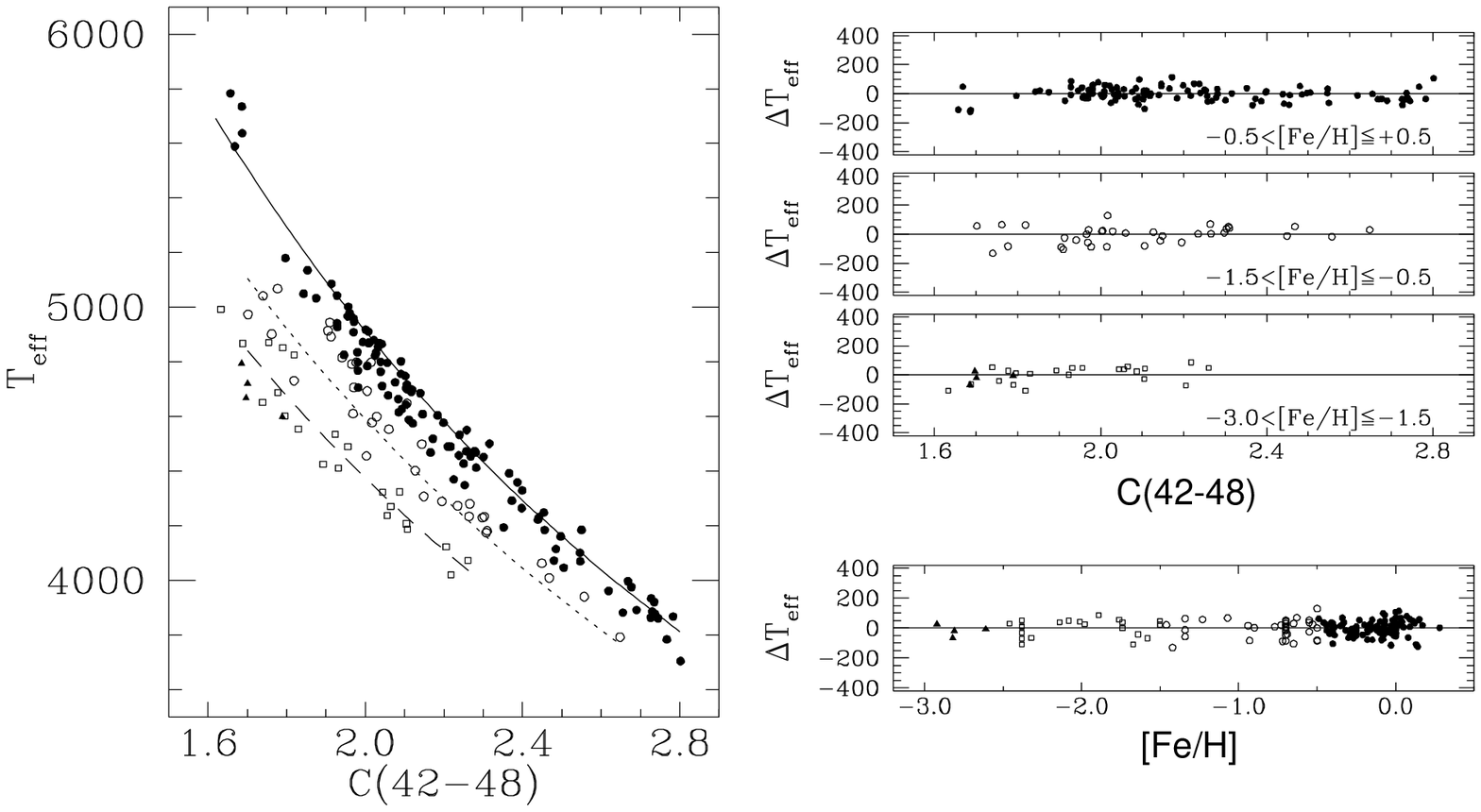}
 \caption{As in Fig. \ref{fig:vilvs} for
          $\teff$ vs $C(42-48)$.}
 \label{fig:ddo4248}
\end{figure*}

For the $C(42-48)$ color index, we found:
\begin{eqnarray}
\thetaeff & = & 0.286+0.370[C(42-48)]-0.081\feh \nonumber \\
& & -0.009\feh^{2} \label{eq:ddo4248}
\end{eqnarray}
with $\sigma(\teff)=50$ K and $N=174$. Since the
$\theta_\mathrm{eff}$ vs $C(42-48)$ relation is almost linear (for
a given $\feh$) the residuals of this original fit do not show any
systematic tendency.

Figure \ref{fig:ddo4248} shows the sample and residuals of this
last fit, whose ranges of applicability are:
\begin{eqnarray}
+1.62<C(42-48)<+2.80 & \textrm{ for} & -0.5<\feh<+0.5, \nonumber\\
+1.70<C(42-48)<+2.64 & \textrm{ for} & -1.5<\feh<-0.5, \nonumber\\
+1.70<C(42-48)<+2.27 & \textrm{ for} & -2.5<\feh<-1.5, \nonumber\\
+1.68<C(42-48)<+1.80 & \textrm{ for} & -3.0<\feh<-2.5. \nonumber
\end{eqnarray}

As it is shown in Fig. \ref{fig:ddo4248}, the $\teff$ vs
$C(42-48)$ lines for constant $\feh$ are almost parallel and thus
the gradient $\Delta\teff/\Delta C(42-48)$ is nearly independent
of $\feh$. Their dependence on color is only slight, going from
approximately 20 K per 0.01 mag at $C(42-48)=1.80$ to 10 K per
0.01 mag at $C(42-48)=2.70$. Due to the strong influence of $\feh$
on this color index, the values of $\Delta\teff/\Delta\feh$ for
Eq. (\ref{eq:ddo4248}) are very high. At $C(42-48)=1.80$ the mean
variation $\Delta\teff/\Delta\feh$ amounts to 135 K per 0.3 dex
for solar metallicity stars and gradually decreases to 60 K per
0.3 dex at $\feh=-2$. At $C(42-48)=2.20$ these quantities reduce
to 100 K per 0.3 dex and 50 K per 0.3 dex, respectively.

As stated previously, it is interesting to note the small values
of the residuals of this fit, which are shown in the right panels
of Fig. \ref{fig:ddo4248}. They confirm that there are no
systematic tendencies introduced by Eq. (\ref{eq:ddo4248}) with
color or metallicity and that the large dispersion in the $\teff$
vs $C(42-48)$ relation can be attributed to well defined strong
metallicity effects. They are also a consequence of an extremely
careful photometric work.

\begin{table*}
\centering
\begin{tabular}{ccrrrrrrrc}\hline\hline
Color ($X$) & Metallicity range & $P_6$ & $P_5$ & $P_4$ & $P_3$ &
$P_2$ & $P_1$ & $P_0$ & Eq.\\ \hline
$(V-S)$ & $-0.5<\feh<+0.5$ &  & 41913 & $-139375$ & 173012 & $-98788$ & 25624 & $-2456.1$ & \ref{eq:vs} \\
$(V-S)$ & $-1.5<\feh<-0.5$ &  & & $-3693$ & 13603 & $-18461$ & 11053 & $-2410.6$ & \ref{eq:vs} \\
$(Y-V)$ & $-0.5<\feh<+0.5$ & $-60268$ & 259060 & $-442050$ & 379138 & $-169791$ & 36864 & $-2962.2$ & \ref{eq:yv} \\
$(Y-V)$ & $-1.5<\feh<-0.5$ & & & & 2751.2 & $-7813.5$ & 7405.1 & $-2267.6$ & \ref{eq:yv} \\
$(Y-V)$ & $-3.0<\feh<-2.5$ & & & & & & 294.14 & $-129.34$ & \ref{eq:yv} \\
$(B_2-V_1)$ & $-0.5<\feh<+0.5$ & $-229.3$ & 1480.6 & $-2382.8$ & 884.94 & 593.81 & $-394.36$ & $29.628$ & \ref{eq:genb2v1} \\
$(B_2-V_1)$ & $-1.5<\feh<-0.5$ & & & & & $-604.77$ & 845.68 & $-224.7$ & \ref{eq:genb2v1} \\
$(B_2-G)$ & $-0.5<\feh<+0.5$ & & & 422.14 & $-563.89$ & $-88.665$ & 216.68 & $-24.021$ & \ref{eq:genb2g} \\
$(B_2-G)$ & $-1.5<\feh<-0.5$ & & & & 66.045 & $-579.78$ & 515.79 & $-30.683$ & \ref{eq:genb2g} \\
$t$ & $-0.5<\feh<+0.5$ & & 3273.6 & $-3220$ & $-1575.9$ & 1879 & $-127.21$ & $-91.596$ & \ref{eq:gent} \\
$t$ & $-1.5<\feh<-0.5$ & & & 13196 & $-21982$ & 11378 & $-1776.3$ & 7.6382 & \ref{eq:gent} \\
$(V-I)_\mathrm{(C)}$ & $-0.5<\feh<+0.5$ & & & 497.01 & $-2572.8$ & 4674.7 & $-3470.1$ & 862.27 & \ref{eq:ricvi} \\
$(V-I)_\mathrm{(C)}$ & $-3.0<\feh<-0.5$ & & & & & & $-86.56$ & 104.53 & \ref{eq:ricvi} \\
\hline
\end{tabular}
\caption{Polynomial fits to the original residuals. The general
form is: $P=P_6X^6+\ldots+P_2X^2+P_1X+P_0$, where the $P_i$'s are
constants and $X$ is the color. The last column specifies the
equation that is being corrected.} \label{table:polfits}
\end{table*}

\subsection{Application to Arcturus}

From the practical point of view, our calibrations allow one to
obtain the effective temperature of a star from its color indices
and atmospheric parameters $\logg$ (a distinction between dwarf
and giant is enough) and $\feh$. Arcturus (also HD 124897) is a
well studied giant star, for which a recent \textit{direct}
effective temperature determination exists (Griffin \& Lynas-Gray
\cite{griffin99}). They derived $\teff=4290\pm30$ K. Our
calibration formulae provide the temperatures listed in Table
\ref{table:arcturus}. Their mean value is $4296\pm28$ K, in
excellent agreement with Griffin \& Lynas-Gray result. It is worth
remarking that this mean value is closer to the direct $\teff$ of
Arcturus than the temperature derived through the IRFM:
$4233\pm55$ K (Alonso et~al. \cite{alonso99a}). This is a direct
consequence of averaging stellar properties from large samples.

\begin{table}
\centering
\begin{tabular}{lcc}\hline\hline
Color & Observed value & $\teff$ (K) \\ \hline
$(V-S)$ & 0.870 & 4253 \\
$(Y-V)$ & 0.910 & 4281 \\
$(B_2-V_1)$ & 0.882 & 4312 \\
$(B_2-G)$ & 0.764 & 4312 \\
$t$ & 0.627 & 4297 \\
$(R-I)_\mathrm{(C)}$ & 0.581 & 4346 \\
$(V-I)_\mathrm{(C)}$ & 1.224 & 4296 \\
$C(42-48)$ & 2.303 & 4271 \\ \hline
\end{tabular}
\caption{The effective temperature of Arcturus according to our
calibrations. Different values have been derived through Eqs.
(\ref{eq:vs})-(\ref{eq:ddo4248}) considering $\feh=-0.54$
(Mel\'endez et~al. \cite{melendez03j}).} \label{table:arcturus}
\end{table}

\section{Comparison with other calibrations} \label{secc:...comp}

\subsection{Vilnius system}

A good reference for this photometric system is Strai\v{z}ys
(\cite{straizys95}) book. He has compiled several works on the
intrinsic colors of stars of several spectral types in the Vilnius
system and has smoothed them to build a single set of colors. This
system is also described in Strai\v{z}ys \& Sviderskiene
(\cite{straizys72}).

\begin{figure*}
 \centering
 \includegraphics[bb=18 440 567 691,width=14cm]{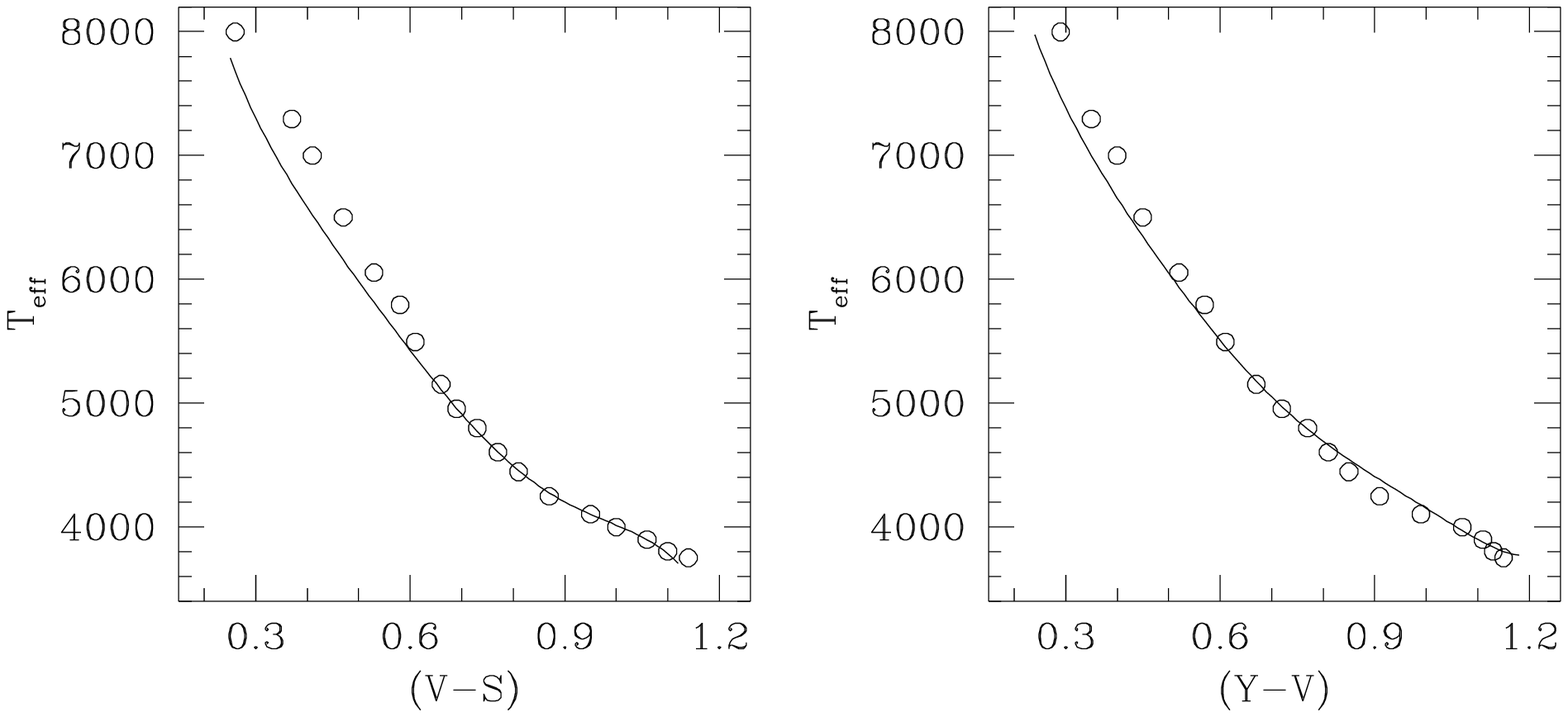}
 \caption{Comparison of our calibrations for the Vilnius colors (solid lines)
 with those given by Strai\v{z}ys (\cite{straizys95}) (open circles).}
 \label{fig:vilcomp}
\end{figure*}

Open circles in Fig. \ref{fig:vilcomp} show the intrinsic $(V-S)$
and $(Y-V)$ colors for giant stars according to Strai\v{z}ys
(\cite[p. 440]{straizys95}), our calibrations for solar
metallicity stars are also shown as solid lines. It is clear that
for temperatures hotter than 5500 K, Strai\v{z}ys calibration
predicts redder colors, the effect being larger for $(V-S)$. The
difference can be as high as 400~K (6\%) at $(V-S)\simeq0.40$ or
500 K (6.7\%) at $(Y-V)\simeq0.30$ with no smooth tendencies.
Below 5200~K the mean difference for $(Y-V)$ is about 1\% and
hardly exceeds 2\%. The agreement for $(V-S)$ in this last range
is remarkable.

The observed difference is partially explained by a gravity
effect, as some of the hottest stars in Fig. \ref{fig:vilvs} are
supergiants. As a matter of fact, the lower a $\logg$ value is, a
bluer $(V-S)$ is obtained (see Sect. \ref{sect:gravityeffects}).

\subsection{Geneva system}

\begin{figure}
 \includegraphics[bb=18 436 282 688,width=7cm]{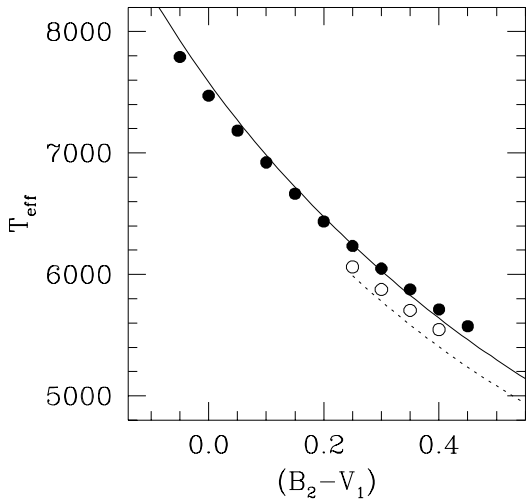}
 \caption{Comparison of our calibration for $\feh=0$ (solid line) and $\feh=-1$
 (dotted line) with Kobi \& North (\cite{kobi90}) $(B_2-V_1)$ colors for
 $\feh=0$ (filled circles) and $\feh=-1$ (open circles).}
 \label{fig:gencomp}
\end{figure}

Kobi \& North (\cite{kobi90}) have published grids of $(B_2-V_1)$
colors based on Kurucz models. A standard correction procedure was
employed to put the synthetic colors into the observational system
though only by using solar metallicity stars. Thus, as remarked by
themselves, their results for metal-poor stars are only reliable
if ``Kurucz models correctly predict the \textit{differential}
effects of blanketing''. In Fig. \ref{fig:gencomp} we compare our
results for $\feh=0$ (solid line) and $\feh=-1$ (dashed line) with
those of Kobi \& North (filled circles correspond to the $\feh=0$
grid while open circles correspond to the $\feh=-1$ grid).

The slopes $\Delta\teff/\Delta(B_2-V_1)$ are slightly different in
Fig.~\ref{fig:gencomp}, the sensitivity of $(B_2-V_1)$ to the
stellar effective temperature is slightly stronger in our
calibration, which implies that Kobi \& North colors result bluer
in the high temperature range and redder below 6000 K. In the
range 6000 K$<\teff<7000$ K there is good agreement.

It is also important to check the effect of $\feh$ on $(B_2-V_1)$
colors according to Kobi \& North. Their colors are such that for
a fixed $(B_2-V_1)$, a change from $\feh=0$ to $\feh=-1$ reduces
the effective temperature by about 170 K, independently of
$(B_2-V_1)$ in the range $0.25<(B_2-V_1)<0.40$. Our calibration
predicts a greater decrease, which amounts from 280 K to 190 K
depending on the value of $(B_2-V_1)$.

\subsection{$RI_\mathrm{(C)}$ system}

\begin{figure*}
 \centering
 \includegraphics[bb=18 440 567 691,width=14cm]{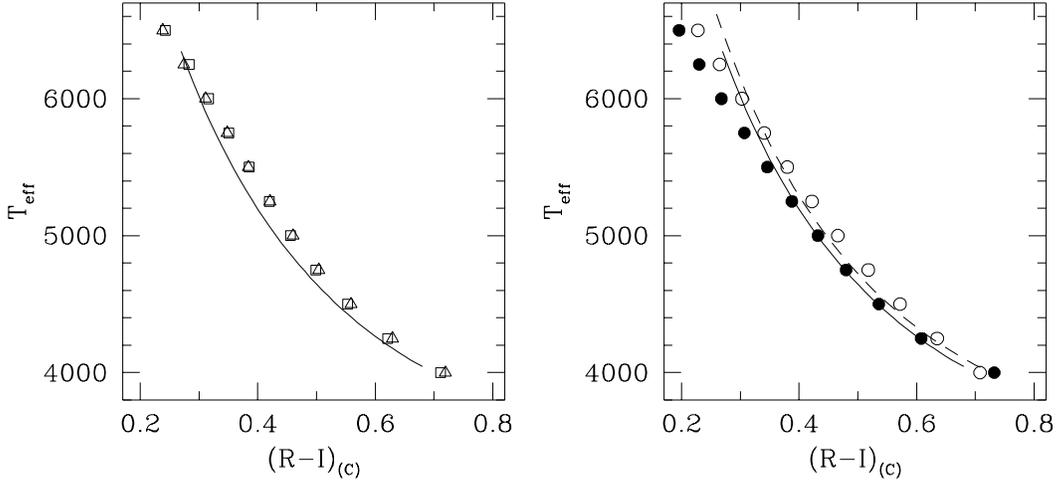}
 \caption{Comparison of our calibration for $(R-I)_\mathrm{(C)}$
 for $\feh=0$ (solid lines) and $\feh=-2$ (dashed line) with the
 $(R-I)_\mathrm{(C)}$ colors obtained by: Bessell et~al. (\cite{bessell98})
 from ATLAS 9 overshoot models (squares) and ATLAS 9 no-overshoot models
 (triangles), Houdashelt et~al. (\cite{houdashelt00}) for $\feh=0$ (filled circles)
 and $\feh=-2$ (open circles).}
 \label{fig:ricomp}
\end{figure*}

\begin{figure}
 \includegraphics[bb=18 436 282 688,width=7cm]{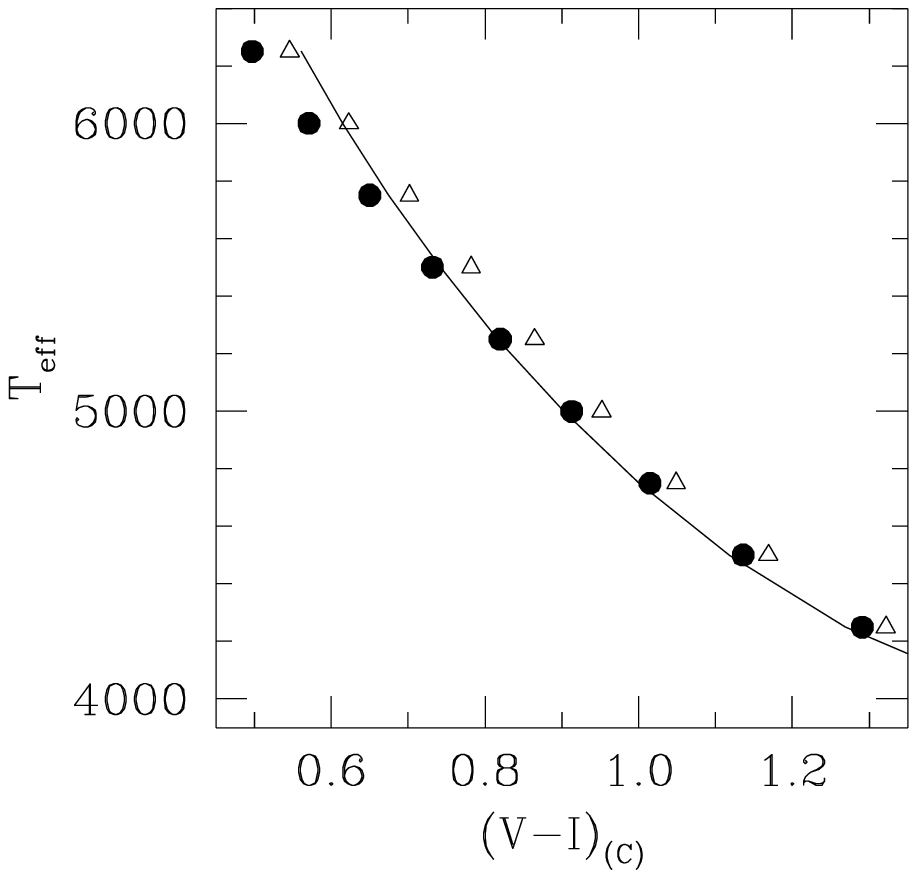}
 \caption{As in Fig. \ref{fig:ricomp} for $(V-I)_\mathrm{(C)}$.}
 \label{fig:vicomp}
\end{figure}

Figures \ref{fig:ricomp} and \ref{fig:vicomp} show our
calibrations for the $\teff$ vs $(R-I)_\mathrm{(C)}$ and $\teff$
vs $(V-I)_\mathrm{(C)}$ relations along with Bessell et~al.
(\cite{bessell98}) and Houdashelt et~al. (\cite{houdashelt00})
results. The former of these provide colors obtained from ATLAS 9
overshoot models (squares) and no-overshoot models (triangles),
which within our interval of interest differ by no more than 0.02
mag. Bessell et~al. colors are valid for solar metallicity stars.
On the other hand, Houdashelt et~al. used improved MARCS models to
calculate colors taking into account metallicity effects. Their
results for $\feh=0$ (filled circles) and $\feh=-2$ (open circles)
are shown. Both Bessell et~al. and Houdashelt et~al. provide
colors for several $\logg$ values, Figs. \ref{fig:ricomp} and
\ref{fig:vicomp} show their results for $\logg=2.0$, the nearest
value to the peak of the $\logg$ distribution of the sample.

The $(R-I)_\mathrm{(C)}$ colors obtained from ATLAS 9 overshoot
models agree slightly better than no-overshoot ones with ours
below $\teff=5500$ K though Bessell et~al. give too red colors in
that range ($\sim0.02$ mag). In the range 5750 K$<\teff<6250$ K,
where no-overshoot models seem to be more reliable (see also
Castelli et~al. \cite{castelli97}) the agreement increases with
temperature. In the case of the $(V-I)_\mathrm{(C)}$ color, below
$\teff=5500$ K there is an almost constant difference of about
0.04 mag that makes our colors (for $\feh=0$ stars) slightly bluer
compared to both overshoot and no-overshoot model colors (only
no-overshoot results are plotted).

Houdashelt et~al. $(R-I)_\mathrm{(C)}$ colors are too blue for
$\teff>5500$~K. The effect gradually increases as hotter stars are
considered. At $\teff=5500$ K, for instance, the difference
amounts to 0.015 mag while at $\teff=6250$ K it is about 0.050
mag. A similar behaviour was also present in the dwarf calibration
(see our Fig. 12c in Paper I). Below $\teff=5000$ K, there is good
agreement since both the gradient
$\Delta\teff/\Delta(R-I)_\mathrm{(C)}$ and $(R-I)_\mathrm{(C)}$
colors are very similar. Even though Houdashelt et~al. colors were
put into the observational system by means of a sample of solar
metallicity stars and thus their results for metal-poor stars
could be not very accurate, the fact that metal-poor stars are
redder than solar metallicity stars for the $(R-I)_\mathrm{(C)}$
color index is very well reproduced and agrees reasonably well
with our result. Nevertheless, according to our calibration, for a
fixed temperature, the difference in $(R-I)_\mathrm{(C)}$ between
a solar metallicity star and a $\feh=-1$ star is no more than 0.02
mag while Houdashelt et~al. give 0.04 mag. On the other hand,
their $(V-I)_\mathrm{(C)}$ colors for solar metallicity stars are
very close to ours, except at high temperatures.

\subsection{DDO system}

\begin{figure*}
 \centering
 \includegraphics[bb=18 440 567 691,width=14cm]{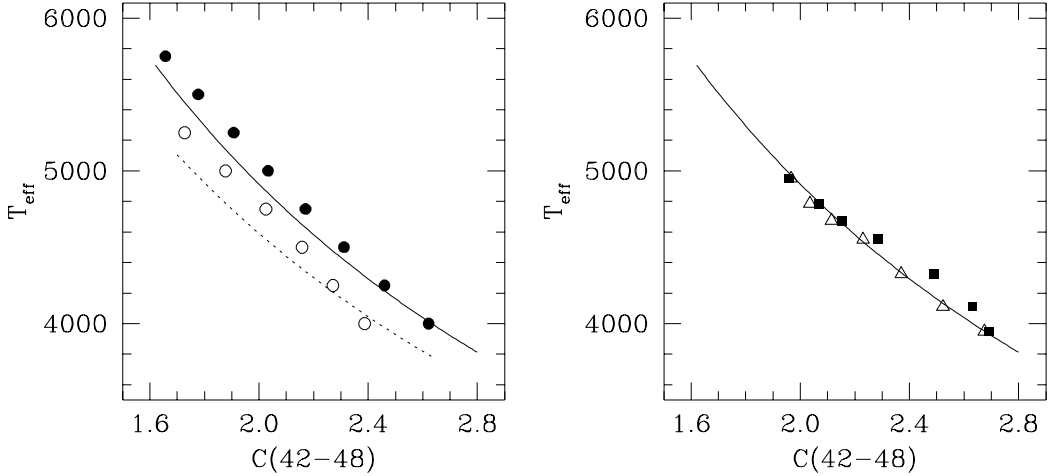}
 \caption{Comparison of our calibration for $C(42-48)$
 for $\feh=0$ (solid line) and $\feh=-1$ (dashed line) with:
 Tripicco \& Bell (\cite{tripicco91}) calibrations for $\feh=0$
 (filled circles) and $\feh=-1$ (open circles), Clari\'a et~al.
 (\cite{claria94}) colors (triangles), and Morossi et~al.
 (\cite{morossi95}) colors (squares).}
 \label{fig:ddocomp}
\end{figure*}

Filled and open circles in Fig. \ref{fig:ddocomp} correspond to
DDO colors for $\feh=0$ and $\feh=-1$ derived from synthetic MARCS
spectra by Tripicco \& Bell (\cite{tripicco91}). Their procedure
to put the synthetic colors into the observational system involves
the use of spectrophotometric scans, a procedure that allows a
model independent treatment. For stars having $\teff>4500$ K,
Tripicco \& Bell colors are too red but the effect of $\feh$ is
well reproduced. Particularly, in the range 4500 K$<\teff<5000$ K,
$C(42-48)$ varies 0.15 mag for a 1.0 dex variation in $\feh$, in
reasonable agreement with the mean 0.12 mag variation that Eq.
(\ref{eq:ddo4248}) produces. Contrary to our calibration, in which
$\teff$ vs $C(42-48)$ lines of constant $\feh$ are almost
parallel, Tripicco \& Bell results suggest stronger metallicity
effects for stars hotter than 5000 K and cooler than 4500 K. In
general, the slopes $\Delta\teff/\Delta C(42-48)$ are quite
different, specially for $\feh=-1$.

The empirical calibration of Clari\'a et~al. (\cite{claria94}) for
the $\teff$ vs $C(42-48)$ relation is shown in Fig.
\ref{fig:ddocomp} with open triangles. They used the
$C(42-45):C(45-48)$ color-color diagram to derive mean DDO colors
for different spectral types, for which a $\teff$ calibration was
also derived. The agreement with our work is quite good, specially
for $\teff<4600$ K. In the range 4600 K$<\teff<5000$ K there is a
shift of only 0.04 mag that makes Clari\'a et~al. colors slightly
bluer than ours.

Finally, Morossi et~al. (\cite{morossi95}) have used Kurucz models
to derive synthetic DDO colors for solar metallicity stars as a
function of spectral type. The mean DDO effective temperature-MK
spectral type provided by Clari\'a et~al. (\cite{claria94}) was
adopted here to derive the $\teff$ vs $C(42-48)$ relation
corresponding to Morossi et~al. colors. This relation provides too
red colors in the range 4100 K$<\teff<4600$~K but agrees with our
color for $\teff=3950$ K. The slope $\Delta\teff/\Delta C(42-48)$
changes dramatically with color, so that around $\teff=4700$ K the
agreement is good again but for hotter stars Morossi et~al. colors
become slightly bluer.

\section{The effective temperature scale} \label{secc:...escala}

\subsection{Intrinsic colors of giant stars}

Instead of fitting the $\mathrm{color}=\mathrm{color}(\teff,\feh)$
relation to some analytical function from the same sample adopted
to derive the $\teff=\teff(\mathrm{color},\feh)$ relation, as is
sometimes done, we solved Eqs. (\ref{eq:vs})-(\ref{eq:ddo4248})
for the color index as a function of $\teff$ and $\feh$ in order
to derive the intrinsic colors of giant stars. The former
procedure does not guarantee a single correspondence between the
three quantities involved and could produce a temperature scale
inconsistent with the calibration formulae.

\begin{figure}
 \includegraphics[bb=17 160 566 689,width=8.5cm]{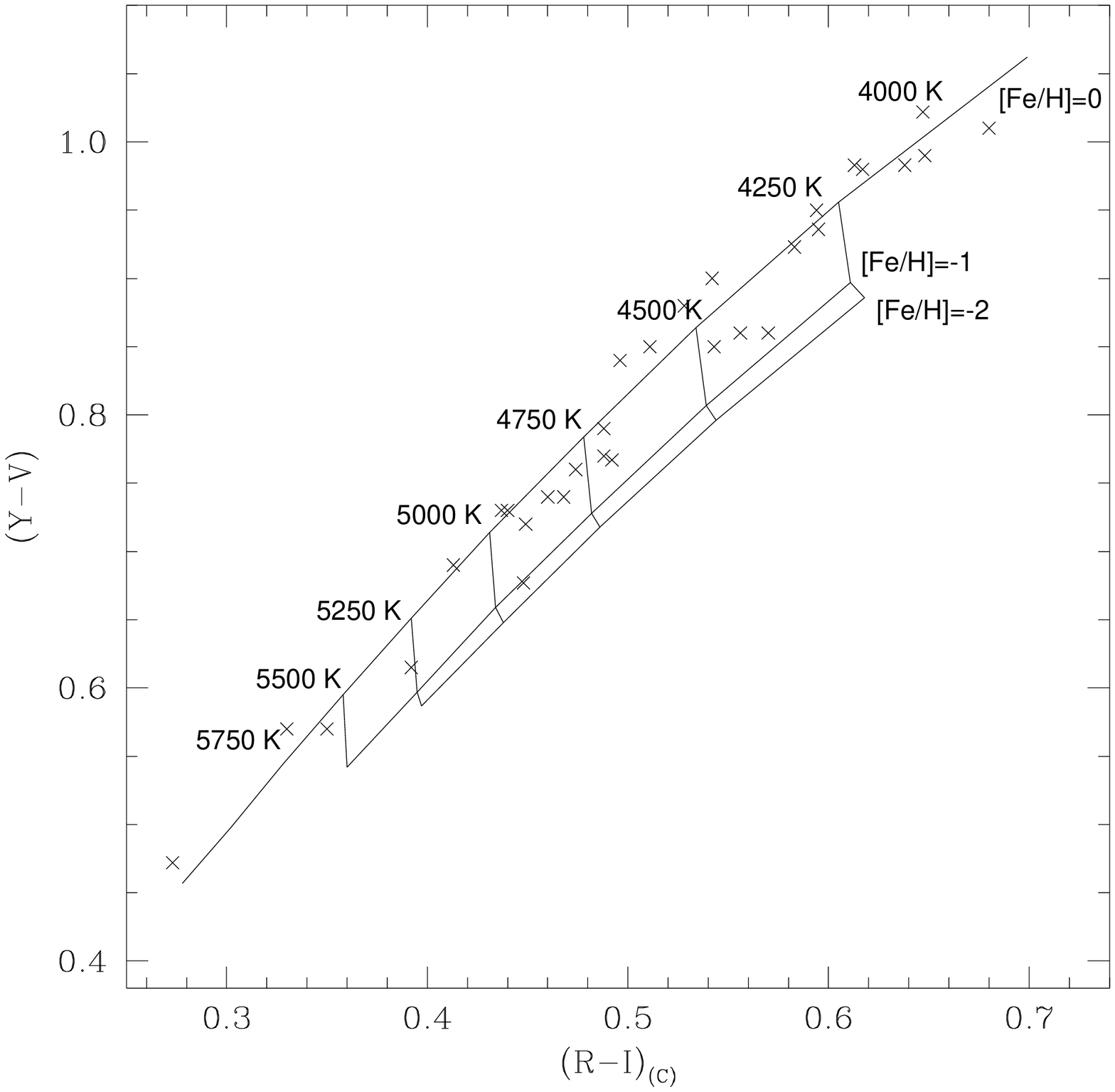}
 \caption{$(R-I)_\mathrm{(C)}$ vs $(Y-V)$, color-color diagram
 showing lines of equal $\teff$ and $\feh$. Crosses correspond to
 derreddened colors of some giant stars with $\feh>-0.5$.}
 \label{fig:rivsyv}
\end{figure}

\begin{figure}
 \includegraphics[bb=17 160 566 689,width=8.5cm]{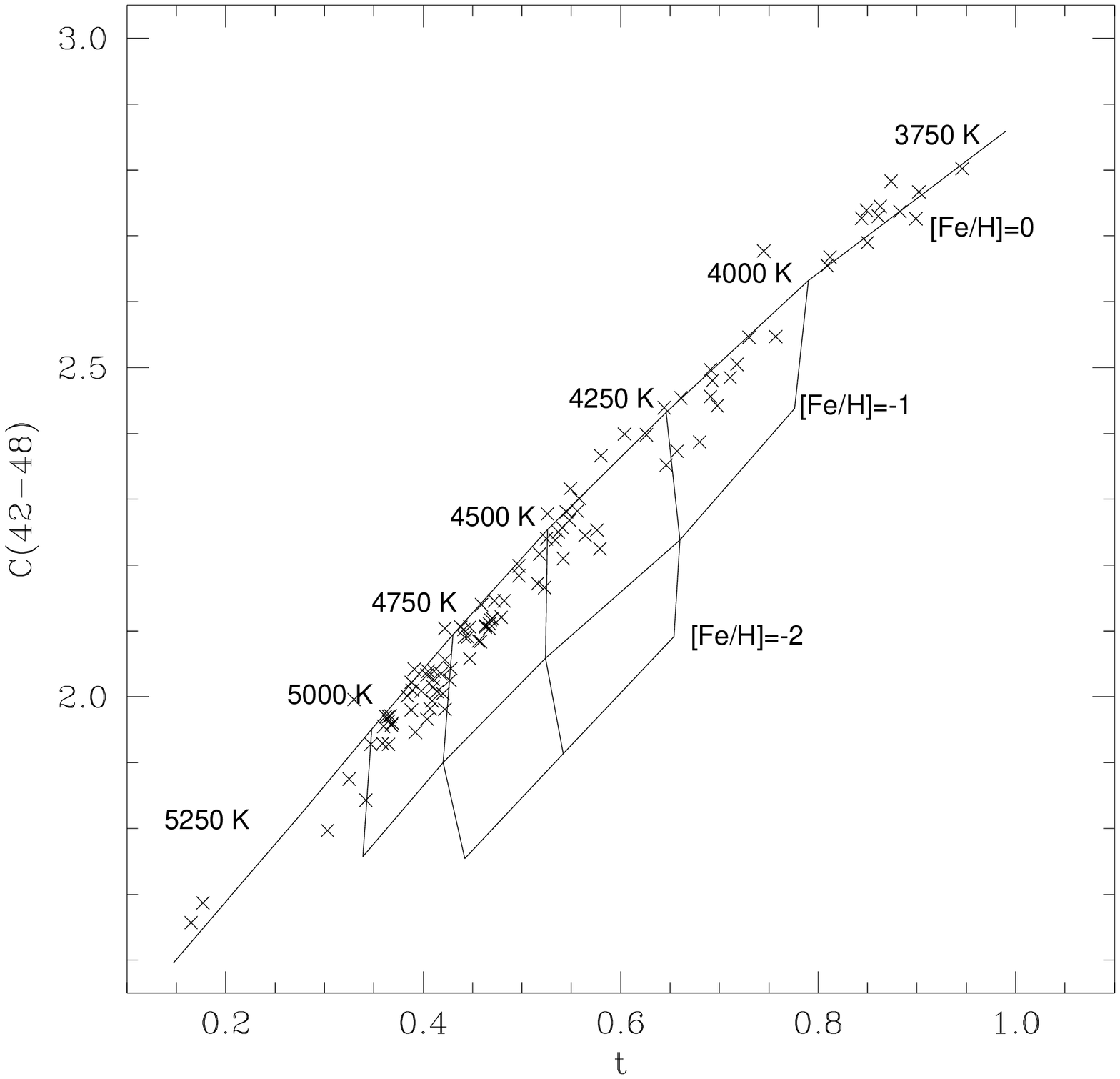}
 \caption{As in Fig. \ref{fig:rivsyv} for $t$ vs $C(42-48)$.}
 \label{fig:tvsc4248}
\end{figure}

Tables \ref{vilnius:intrinsic-colors},
\ref{geneva:intrinsic-colors} and \ref{ric-ddo:intrinsic-colors}
list the intrinsic colors of giant stars as a function of $\teff$
and $\feh$, including some reliable extrapolated values. They have
been used to plot the color-color diagrams of Figs.
\ref{fig:rivsyv} and \ref{fig:tvsc4248}, in which we also show the
derreddened colors of some giant stars with $\feh>-0.5$. These
stars have been introduced to verify that our calibrations
satisfactorily reproduce the empirical color-color diagrams. We
attribute the observed dispersion to the amplitude of the $\feh$
interval covered and to the fact that only a few stars have
$\feh>0$.

\begin{table*}
\centering
\begin{tabular}{c r c c r c c}\hline\hline
 & \multicolumn{3}{c}{$(V-S)$} & \multicolumn{3}{c}{$(Y-V)$} \\ \hline
$\teff$ &   $\feh=0.0$  &   $-1.0$  &   $-2.0$  &   $\feh=0.0$  &   $-1.0$  &   $-2.0$  \\  \hline
 3750    &   1.109   &   -   &   -   &   -   &   -   &   -   \\
 3800    &   1.095   &   -   &   -   &   1.152   &   -   &   -   \\
 4000    &   1.008   &   0.935   &   0.952   &   1.059   &   0.969   &   -   \\
 4250    &   0.880   &   0.853   &   0.864   &   0.961   &   0.870   &   0.886   \\
 4500    &   0.798   &   0.779   &   0.785   &   0.866   &   0.787   &   0.796   \\
 4750    &   0.735   &   0.712   &   0.715   &   0.782   &   0.717   &   0.718   \\
 5000    &   0.682   &   0.653   &   0.652   &   0.713   &   0.657   &   0.648   \\
 5250    &   0.633   &   0.600   &   0.594   &   0.654   &   0.606   &   0.587   \\
 5500    &   0.586   &   0.553   &   -   &   0.602   &   0.560   &   -   \\
 5750    &   0.541   &   0.511   &   -   &   0.554   &   0.520   &   -   \\
 6000    &   0.497   &   -   &   -   &   0.509   &   -   &   -   \\
 6250    &   0.454   &   -   &   -   &   0.466   &   -   &   -   \\
 6500    &   0.413   &   -   &   -   &   0.425   &   -   &   -   \\
 6750    &   0.374   &   -   &   -   &   0.386   &   -   &   -   \\
 7000    &   0.338   &   -   &   -   &   0.350   &   -   &   -   \\
 7250    &   0.307   &   -   &   -   &   0.317   &   -   &   -   \\
 7500    &   0.278   &   -   &   -   &   0.287   &   -   &   -   \\
 7750    &   0.253   &   -   &   -   &   0.261   &   -   &   -   \\  \hline
\end{tabular}
\caption{Intrinsic colors of giant stars in the Vilnius system as
a function of $\feh$.} \label{vilnius:intrinsic-colors}
\end{table*}

\begin{table*}
\centering
\begin{tabular}{c r c c r c c r c c}\hline\hline
 & \multicolumn{3}{c}{$(B_2-V_1)$} & \multicolumn{3}{c}{$(B_2-G)$} & \multicolumn{3}{c}{$t$} \\ \hline
$\teff$ &   $\feh=0.0$  &   $-1.0$  &   $-2.0$  &   $\feh=0.0$  &   $-1.0$  &   $-2.0$  &   $\feh=0.0$  &   $-1.0$  &   $-2.0$  \\ \hline
3500    &   -   &   -   &   -   &   -   &   -   &   -   &   0.990   &   -   &   -   \\
3750    &   1.224   &   -   &   -   &   1.198   &   -   &   -   &   0.909   &   -   &   -   \\
4000    &   1.077   &   1.052   &   -   &   1.028   &   0.984   &   -   &   0.790   &   0.776   &   -   \\
4250    &   0.931   &   0.870   &   0.871   &   0.838   &   0.731   &   0.726   &   0.646   &   0.660   &   0.654   \\
4500    &   0.803   &   0.733   &   0.729   &   0.658   &   0.549   &   0.546   &   0.526   &   0.524   &   0.542   \\
4750    &   0.693   &   0.623   &   0.604   &   0.503   &   0.403   &   0.385   &   0.430   &   0.420   &   0.442   \\
5000    &   0.599   &   0.528   &   0.494   &   0.371   &   0.279   &   0.240   &   0.348   &   0.339   &   0.353   \\
5250    &   0.515   &   0.446   &   0.396   &   0.257   &   0.172   &   0.109   &   0.276   &   0.270   &   0.271   \\
5500    &   0.440   &   0.373   &   0.308   &   0.155   &   0.076   &   $-0.010$  &   0.209   &   0.208   &   0.197   \\
5750    &   0.371   &   0.307   &   -   &   0.064   &   $-0.010$  &   -   &   0.147   &   0.147   &   -   \\
6000    &   0.307   &   0.248   &   -   &   $-0.020$  &   $-0.089$  &   -   &   0.088   &   0.085   &   -   \\
6250    &   0.249   &   -   &   -   &   $-0.097$  &   -   &   -   &   0.031   &   -   &   -   \\
6500    &   0.194   &   -   &   -   &   $-0.170$  &   -   &   -   &   $-0.023$  &   -   &   -   \\
6750    &   0.144   &   -   &   -   &   $-0.239$  &   -   &   -   &   $-0.076$  &   -   &   -   \\
7000    &   0.097   &   -   &   -   &   $-0.303$  &   -   &   -   &   $-0.126$  &   -   &   -   \\
7250    &   0.053   &   -   &   -   &   $-0.365$  &   -   &   -   &   $-0.173$  &   -   &   -   \\
7500    &   0.013   &   -   &   -   &   $-0.425$  &   -   &   -   &   $-0.218$  &   -   &   -   \\
7750    &   $-0.025$  &   -   &   -   &   $-0.481$  &   -   &   -   &   $-0.260$  &   -   &   -   \\
8000    &   $-0.060$  &   -   &   -   &   $-0.536$  &   -   &   -   &   $-0.299$  &   -   &   -   \\
8250    &   $-0.093$  &   -   &   -   &   $-0.588$  &   -   &   -   &   -   &   -   &   -   \\ \hline
\end{tabular}
\caption{As in Table \ref{vilnius:intrinsic-colors} for the Geneva
system.} \label{geneva:intrinsic-colors}
\end{table*}

\begin{table*}
\centering
\begin{tabular}{c r c c r c c r c c}\hline\hline
 & \multicolumn{3}{c}{$(R-I)_\mathrm{(C)}$} & \multicolumn{3}{c}{$(V-I)_\mathrm{(C)}$} & \multicolumn{3}{c}{$C(42-48)$} \\ \hline
$\teff$ &   $\feh=0.0$  &   $-1.0$  &   $-2.0$  &   $\feh=0.0$  &   $-1.0$  &   $-2.0$  & $\feh=0.0$  &   $-1.0$  &   $-2.0$\\ \hline
3750    &   -   &   -   &   -   &   1.941   &   -   &   -   &   2.859   &   -   &   -   \\
4000    &   0.699   &   -   &   0.718   &   1.491   &   -   &   -   &   2.632   &   2.438   &   -   \\
4250    &   0.605   &   0.611   &   0.618   &   1.269   &   -   &   1.233   &   2.432   &   2.238   &   2.092   \\
4500    &   0.534   &   0.539   &   0.544   &   1.118   &   1.097   &   1.078   &   2.254   &   2.059   &   1.914   \\
4750    &   0.478   &   0.482   &   0.486   &   1.000   &   0.971   &   0.956   &   2.095   &   1.900   &   1.754   \\
5000    &   0.431   &   0.434   &   0.438   &   0.902   &   0.869   &   -   &   1.951   &   1.757   &   -   \\
5250    &   0.392   &   0.395   &   0.397   &   0.817   &   -   &   -   &   1.822   &   -   &   -   \\
5500    &   0.358   &   0.360   &   0.363   &   0.742   &   -   &   -   &   1.704   &   -   &   -   \\
5750    &   0.328   &   0.330   &   0.332   &   0.675   &   -   &   -   &   1.596   &   -   &   -   \\
6000    &   0.302   &   0.304   &   0.306   &   0.615   &   -   &   -   &   -   &   -   &   -   \\
6250    &   0.278   &   0.280   &   0.282   &   0.561   &   -   &   -   &   -   &   -   &   -   \\ \hline
\end{tabular}
\caption{As in Table \ref{vilnius:intrinsic-colors} for the
$RI_\mathrm{(C)}$ and DDO systems.}
\label{ric-ddo:intrinsic-colors}
\end{table*}

\subsection{Gravity effects} \label{sect:gravityeffects}

\begin{figure*}
 \centering
 \includegraphics[bb=19 9 565 177,width=15cm]{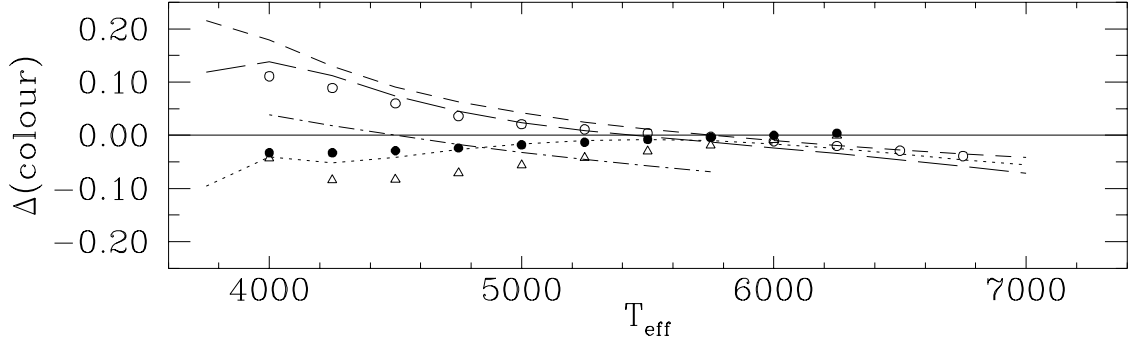}
 \caption{Gravity effects on photometric color indices. The vertical
 axis measures the difference in color between a solar metallicity
 giant and a main sequence star as a function of $\teff$.
 A giant is redder than a main sequence star of the same effective
 temperature when $\Delta$(color) is positive. Colors plotted
 are: $(V-S)$ (dotted line), $(Y-V)$ (open circles), $(B_2-V_1)$
 (dashed line), $(B_2-G)$ (long-dashed line), $(R-I)_\mathrm{(C)}$
 (filled circles), $(V-I)_\mathrm{(C)}$ (triangles), and $C(42-48)$ (dash-dotted line).}
 \label{fig:gravity}
\end{figure*}

Colors derived in this work for giants generally differ from those
given for main sequence stars in Paper I. Even though the mean
differences are of the order of the observational errors, so that
they are not very useful when studying individual stars, knowledge
of gravity effects on the empirical temperature scale may improve
our understanding of stellar spectra. The difference in color
between a giant and a dwarf star of the same $\teff$ is plotted in
Fig.~\ref{fig:gravity} for stars with $\feh=0$. Different symbols
and line styles correspond to different color indices.

Firstly, it is worth noticing that $(R-I)_\mathrm{(C)}$ is almost
unaffected by the surface gravity, dwarfs are only 0.03 mag redder
than giants at $\teff=4000$ K and non-distinguishable from them at
high temperatures. Given that it is also nearly independent of
$\feh$, this color is a good $\teff$ indicator, free of secondary
effects.

For most colors, main sequence F stars are redder than F giants.
The difference seems to increase with the separation between the
wavelength bands covered by each filter. It is large (0.07 mag at
$\teff=7000$ K) for $(B_2-G)$, whose $B_2$ filter has a mean
wavelength that is about 1300\AA\ away from the $G$ one. It is
relatively large (0.06 mag) for $(V-S)$, for which the separation
is approximately 1100\AA, and less prominent (0.04 mag) for
$(Y-V)$ and $(B_2-V_1)$, whose separations are around 850\AA.
Neutral hydrogen bound-free opacity increases more with wavelength
than that of the H$^{-}$ ion, which is the only of the two that
increases with electron pressure, and so is stronger in main
sequence stars. This implies that the wavelength dependence of the
total opacity is smoother for dwarfs and stronger for giants. As a
result, within Paschen continuum (365--830 nm), giants gradually
emit more radiation at shorter wavelengths than dwarfs, an effect
that is well reproduced in Kurucz SEDs. Therefore, in addition to
their primary dependence on $\teff$, colors at well separated
wavelength intervals within Paschen continuum will be more
affected by the surface gravity. Alonso et~al. (\cite{alonso99b})
also showed this effect for $(B-V)$ and $(V-I)$ and provided a
similar explanation.

On the other hand, UV and optical colors of K stars are affected
by different mechanisms, which can make a giant bluer or redder
than a main sequence star. In addition to their dependence with
the continuum opacity, at the low $\teff$s found in K stars,
stellar spectra are crowded with lots of lines that can be
stronger in giants (e.g. CH lines) or in dwarfs (e.g. MgH lines),
as explained by Tripicco \& Bell (\cite{tripicco91}) and Paltoglou
\& Bell (\cite{paltoglou94}).

Though it is not shown here, we have also checked that colors for
metal-poor stars are only slightly affected by the surface
gravity.

\subsection{Giants in clusters}

\begin{figure}
 \includegraphics[bb=19 10 290 309,width=8.6cm]{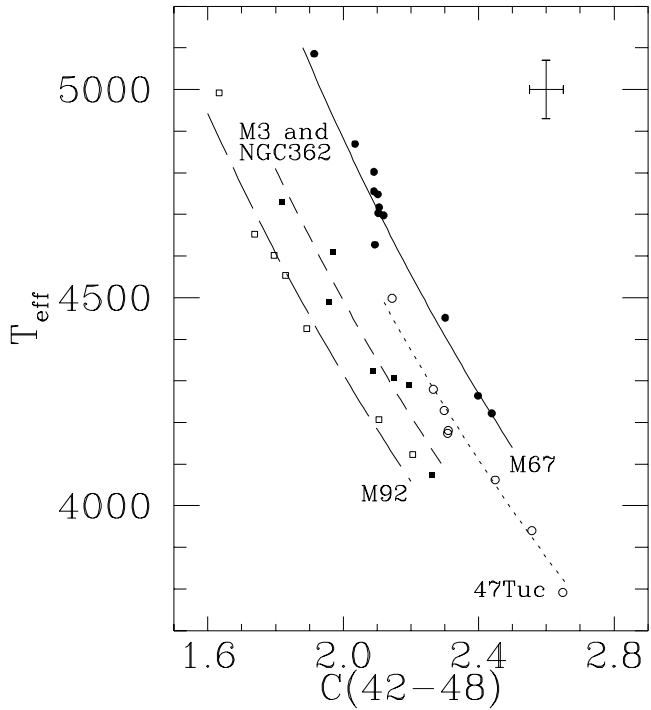}
 \caption{$\teff$ vs $C(42-48)$ relations for giants in the open
 cluster M67 (filled circles) and in the globular clusters 47 Tuc
 (open circles), NGC 362 and M3 (filled squares), and M92 (open
 squares). Solid, dotted, dashed and long-dashed lines correspond to our
 calibrations for $\feh=-0.08$, $\feh=-0.70$, $\feh=-1.50$ and
 $\feh=-2.38$, respectively, which are the mean metallicities of the
 clusters. Error bars are shown to the upper right corner of this
 figure.}
 \label{fig:escalag}
\end{figure}

The effective temperature scale derived in this work reproduces
well the $\teff$ versus color relations of open and old globular
clusters, as is shown in Fig. \ref{fig:escalag} for the $C(42-48)$
color. Table \ref{table:ddo-clusters} contains the temperatures of
the stars in this figure as given by Alonso et~al.
(\cite{alonso99b}) and intrinsic colors from the GCPD corrected by
using the $E(B-V)$ values given by KI03. The observed deviations
are within the observational errors, which can be verified from
the error bars to the upper right corner of Fig.
\ref{fig:escalag}. From this result not only we conclude that the
temperature scale derived here is suited for stars in clusters,
but also we are showing that the metallicity and reddening scales
for these clusters (KI03) are well determined.

\begin{table*}
\centering
\begin{tabular}{ccc|ccc|ccc} \hline\hline
Star & $\teff$ & $C(42-48)$ & Star & $\teff$ & $C(42-48)$ & Star & $\teff$ & $C(42-48)$ \\ \hline
M3 III28    &   4073    &   2.261   &   M67 231 &   4869    &   2.035   &   47Tuc 4418  &   3940    &   2.557   \\
M3 IV25 &   4324    &   2.087   &   M67 244 &   5086    &   1.914   &   47Tuc 5406  &   4181    &   2.310   \\
M3 216  &   4490    &   1.956   &   M67 I17 &   4933    &   1.996   &   47Tuc 5427  &   4229    &   2.298   \\
M67 84  &   4748    &   2.102   &   M67 IV20    &   4627    &   2.093   &   47Tuc 5527  &   4494    &   2.043   \\
M67 105 &   4452    &   2.301   &   M92 III13   &   4123    &   2.206   &   47Tuc 5529  &   3792    &   2.648   \\
M67 108 &   4222    &   2.439   &   M92 VII18   &   4207    &   2.105   &   47Tuc 5627  &   4174    &   2.308   \\
M67 141 &   4755    &   2.091   &   M92 XII8    &   4425    &   1.893   &   47Tuc 5739  &   4062    &   2.449   \\
M67 151 &   4802    &   2.091   &   M92 IV10    &   4553    &   1.830   &   47Tuc 6509  &   4498    &   2.144   \\
M67 164 &   4698    &   2.119   &   M92 IV2 &   4602    &   1.796   &   NGC362 I44  &   4289    &   2.195   \\
M67 170 &   4264    &   2.398   &   M92 IV114   &   4652    &   1.739   &   NGC362 II43 &   4610    &   1.969   \\
M67 223 &   4717    &   2.106   &   M92 III4    &   4992    &   1.633   &   NGC362 II47 &   4731    &   1.819   \\
M67 224 &   4703    &   2.104   &   47Tuc 3407  &   4280    &   2.266   &   NGC362 III4 &   4307    &   2.149   \\ \hline
\end{tabular}
\caption{Temperatures and (intrinsic) colors for the stars plotted in Fig.
\ref{fig:escalag}.} \label{table:ddo-clusters}
\end{table*}

\subsection{Evolutionary calculations}

One of the most important applications of the temperature scale is
the transformation of theoretical HR diagrams into CMDs since it
allows to explore the capability of evolutionary calculations to
reproduce the observations.

\begin{figure*}
\includegraphics[bb=34 9 565 311,width=17cm]{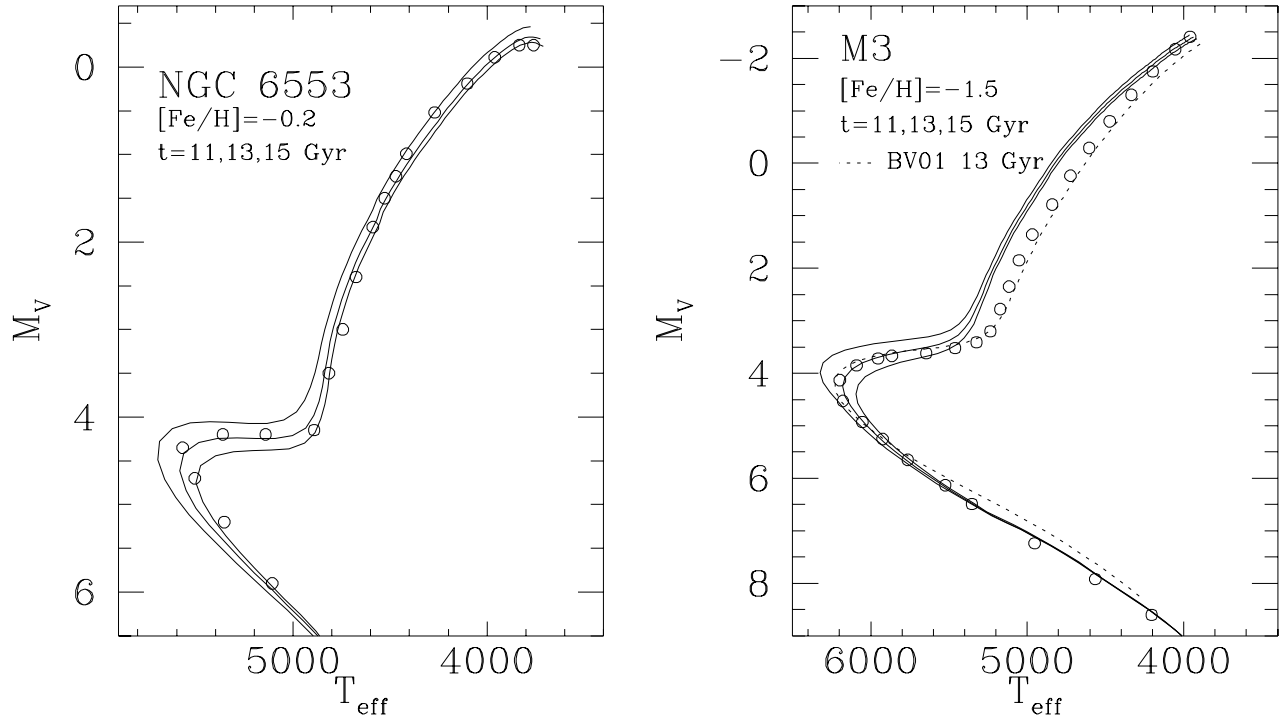}
\caption{Left: Fiducial line for NGC 6553 according to Guarnieri
et~al. (\cite{guarnieri98}) and Ortolani et~al.
(\cite{ortolani95}) transformed to the $M_V$ vs $\teff$ plane from
our calibrations (open circles). The solid lines are theoretical
isochrones calculated by Yi et~al. (\cite{yi03}). Right: As in the
left panel for M3 (Root et~al. \cite{rood99}, Johnson \& Bolte
\cite{johnson98}). The dotted line is an isochrone from Bergbusch
\& VandenBerg (\cite{bergbusch01}, BV01) grids.} \label{fig:iso}
\end{figure*}

Here we compare isochrones with fiducial lines for two globular
clusters: NGC 6553 and M3. Figure \ref{fig:iso} shows these
fiducial lines along with theoretical isochrones for $\feh=-0.2$,
($[\alpha/\mathrm{Fe}]=+0.2$) and $\feh=-1.5$
($[\alpha/\mathrm{Fe}]=+0.3$) and ages $t=11,$ 13 and 15 Gyr;
according to Yi et~al. (\cite{yi03}, Y$^2$). Also shown is the
Bergbusch \& VandenBerg (\cite{bergbusch01}, BV01) isochrone for
$\feh=-1.5$ ($[\alpha/\mathrm{Fe}]=+0.3$) and $t=13$ Gyr.

The bulge globular cluster NGC 6553 serves as a template for
metal-rich galactic populations (ellipticals and bulges) given
that it is one of the most metal-rich globular clusters of the
Galaxy ($\feh=-0.2$, Melendez et~al. \cite{melendez03j}). Data for
this cluster are from Guarnieri et~al. (\cite{guarnieri98}) and
Ortolani et~al. (\cite{ortolani95}). Their HST
$(V-I)_\mathrm{(C)}$ colors have been transformed into $\teff$ by
using our calibrations for both dwarfs (Paper I) and giants. We
adopted $E(V-I)$ and $(m-M)$ from Guarnieri et al. (1998). The
best fit occurs at $t=13$~Gyr, in agreement with the old age
obtained by Ortolani et al. (\cite{ortolani95}).

For the halo globular cluster M3 ($\feh=-1.5$, KI03), almost all
the photometry is from HST (Rood et~al. \cite{rood99}), the last
three points on the main sequence are ground-base observations
(Johnson \& Bolte \cite{johnson98}) corrected for blending using
the HST data. The $E(B-V)$ and $(m-M)$ values were also taken from
KI03. Again, colors were transformed into $\teff$ from our
temperature scale. In order to obtain a reasonable agreement with
the models, M3 photometry has been empirically corrected by
$\Delta\teff=+60$ K and $\Delta(m-M)=+0.2$ to fit the turnoff of
the Y$^2$ isochrones. Likewise, the BV01 isochrone has been
shifted to fit the observed turnoff.

Note that only a small adjustment is required to obtain a better
fit to the data, the $\Delta\teff=+60$ K is equivalent to a
correction of only 0.014 mag in  $(V-I)_\mathrm{(C)}$ (or $\Delta
E(B-V)=0.01$ mag), and the correction for the absolute magnitude
is well between the error bars for the distance modulus. For M3,
$(m-M)$ ranges from 14.8 (Kraft et al. \cite{kraft92}) to 15.2
(BV01). A correction of 0.2 mag in $M_V$ corresponds to a change
of only 0.02 dex in the iron abundance obtained from Fe~II (KI03).
It is important to note that the isochrone of BV01 satisfactorily
reproduces the observed RGB, but the Y$^2$ isochrones fit better
the low main sequence.

\section{Conclusions} \label{secc:...conclusion}

We have calibrated the effective temperature versus color
relations for several color indices in 4 important photometric
systems using reliable and recent $\feh$ and $\logg$ measurements
to explore with improved accuracy the effects of chemical
composition and surface gravity on the temperature scale.

In general, the present calibrations span the following ranges:
3800 K$<\teff<$8000 K, $-3.0<\feh<+0.5$. Ranges of applicability,
however, are different and specific for each color calibration.
For $(R-I)_\mathrm{(C)}$ and $C(42-48)$, for example, these ranges
are not so wide, specially in $\teff$. We also provide specific
ranges of applicability after every formula.

The standard deviation of the fits amount from 46 K for
$(V-I)_\mathrm{(C)}$ to 99 K for $(Y-V)$ with more than 130 stars
in almost every calibration. Residuals of every fit were
iteratively checked to reduce the dispersion and undesirable
systematic effects.

Finally, from our formulae we have calculated the intrinsic colors
of giant stars and have probed their consistency with empirical
color-color diagrams, gravity effects on stellar spectra, $\teff$
versus color relations for stars in clusters and evolutionary
calculations. Our results for the main sequence were also explored
in this last case.

\begin{acknowledgements}
We thank the referee Dr.~M.~Bessell for his comments and
suggestions which have certainly improved this paper. This work
was supported by the Consejo Nacional de Ciencia y Tecnolog\'ia
(CONCYTEC, 253-2003) of Per\'u. I.R. acknowledges support from the
Exchange of Astronomers Programme of the IAU.
\end{acknowledgements}


\begin{thebibliography}{}

\bibitem[1996a]{alonso96a}
  Alonso, A., Arribas, S., \& Mart\'{\i}nez-Roger, C.
  1996a, \aaps, 117, 227

\bibitem[1996b]{alonso96b}
  Alonso, A., Arribas, S., \& Mart\'{\i}nez-Roger, C.
  1996b, \aap, 313, 873

\bibitem[1999a]{alonso99a}
  Alonso, A., Arribas, S., \& Mart\'{\i}nez-Roger, C.
  1999a, \aaps, 139, 335

\bibitem[1999b]{alonso99b}
  Alonso, A., Arribas, S., \& Mart\'{\i}nez-Roger, C.
  1999b, \aaps, 140, 261

\bibitem[1997]{bell97}
  Bell R. A.
  1997, Tests of effective temperature--color relations. In
  Proceedings of IAU Symp. 189, ``Fundamental Stellar Properties: the
  interaction between observation and theory'', ed. T. R. Bedding,
  A. J. Booth, \& J.  Davis. Kluwer Academic Publishers, 159

\bibitem[2001]{bergbusch01}
 Bergbusch, P. A., \& VandenBerg, D. A.
 2001, \apj, 556, 322 (BV01)

\bibitem[1979]{bessell79}
  Bessell, M. S.
  1979, \pasp, 91, 589

\bibitem[1998]{bessell98}
  Bessell, M. S., Castelli, F., \& Plez, B.
  1998, \aap, 333, 231

\bibitem[2001]{bessell01}
  Bessell, M. S.
  2001, \pasp, 113, 66

\bibitem[1997]{castelli97}
  Castelli, F., Gratton, R. G., \& Kurucz, R. L.
  1997, \aap, 318, 841

\bibitem[1992]{cayrel92}
  Cayrel de Strobel, G., Hauck, B., Fran\c{c}ois, T., et~al.
  1992, \aaps, 95, 273

\bibitem[2001]{cayrel01}
  Cayrel de Strobel, G., Soubiran, C., \& Ralite, N.
  2001, \aap, 373, 159

\bibitem[1994]{claria94}
  Clari\'a, J. J., Piatti, A. E., \& Lapasset, E.
  1994, \pasp, 106, 436

\bibitem[1999]{cramer99}
  Cramer, N.
  1999, NewAr, 43, 343

\bibitem[2002]{girardi02}
  Girardi, L., Bertelli, G., Bressan, A., et~al.
  2002, \aap, 391, 195

\bibitem[1999]{griffin99}
  Griffin, R. E. M., \& Lynas-Gray, A. E.
  1999, \aj, 117, 2998

\bibitem[1998]{guarnieri98}
 Guarnieri, M. D., Ortolani, S., Montegriffo, P., et~al.
 1998, \aap, 331, 70

\bibitem[1996]{hauck96}
  Hauck, B. \& K\"{u}nzli, M.
  1996, Baltic Astron., 5, 303

\bibitem[2003]{heiter03}
  Heiter, U., \& Luck, R. E.
  2003, Abundance analysis of planetary host stars.
  In proceedings of IAU Symp. 210, ``Modelling of Stellar Atmospheres",
  ed. N. Piskunov, W.W. Weiss, \& D.F. Gray, ASP Conference
  Series, in press

\bibitem[2000]{hilker00}
  Hilker, M.
  2000, \aap, 355, 994 (H00)

\bibitem[2000]{houdashelt00}
  Houdashelt, M. L., Bell, R. A, \& Sweigart, A. V.
  2000, \aj, 119, 1448

\bibitem[1998]{johnson98}
  Johnson, J. A., Bolte, M.
  1998, \aj, 115, 693

\bibitem[1990]{kobi90}
  Kobi, D., \& North, P.
  1990, \aaps, 85, 999

\bibitem[1992]{kraft92}
  Kraft, R., Sneden, C., Langer, G. E., \& Prosser, C.
  1992, \aj, 104, 645

\bibitem[2003]{kraft03}
  Kraft, R., \& Ivans, I.
  2003, \pasp, 115, 143 (KI03)

\bibitem[1995]{ortolani95}
  Ortolani, S., Renzini, A., Gilmozzi, R., et~al.
  1995, Nature, 377, 701

\bibitem[2002]{melendez02}
  Mel\'endez, J., \& Barbuy, B.
  2002, \apj, 575, 474

\bibitem[2003]{melendez03j}
  Melendez, J., Barbuy, B., Bica, B., et~al.
  2003, \aap, 411, 417

\bibitem[2003]{melendez03}
  Mel\'endez, J., \& Ram\'{\i}rez, I.
  2003, \aap, 398, 705 (Paper I)

\bibitem[1997]{mermilliod97}
  Mermilliod, J. C., Mermilliod, M., \& Hauck, B.
  1997, \aaps, 124, 349 (GCPD)

\bibitem[2001]{mishenina01}
  Mishenina, T. V., \& Kovtyukh V. V.
  2001, \aap, 370, 951

\bibitem[1995]{morossi95}
  Morossi, C., Franchini, M., Malagnini, M. L., \& Kurucz, R. L.
  1995, \aap, 295, 471

\bibitem[2002]{nissen02}
  Nissen, P. E., Primas, F., Asplund, M., \& Lambert, D. L.
  2002, \aap, 390, 235

\bibitem[1994]{paltoglou94}
  Paltoglou, G., \& Bell, R. A.
  1994, \mnras, 268, 793

\bibitem[1999]{rood99}
  Rood, R. T., Carretta, E., Paltrinieri, B., et~al.
  1999, \apj, 523, 752

\bibitem[1999]{ryan99}
  Ryan, S. G., Norris, J. E., \& Beers, T. C.
  1999, \apj, 523, 654

\bibitem[2001]{santos01}
  Santos, N. C., Israelian, G. \& Mayor, M.
  2001, \aap, 373, 1019

\bibitem[2002]{smith02}
  Smith, V., Hinkle, K. H., Cunha, K., et~al.
  2002, \aj, 124, 3241

\bibitem[2002]{stephens02}
  Stephens, A., \& Boesgaard, A. M.
  2002, \aj, 123, 1647

\bibitem[1972]{straizys72}
  Strai\v{z}ys, V., \& Sviderskiene, Z.
  1972, \aap, 17, 312

\bibitem[1995]{straizys95}
  Strai\v{z}ys, V. 1995, in Multicolor Stellar Photometry
  (Pachart Publishing House), 430

\bibitem[2002]{takada02}
  Takada-Hiday, M., Takeda, Y., Sato, S., et~al.
  2002, \apj, 573, 614

\bibitem[1991]{tripicco91}
  Tripicco, M., \& Bell, R. A.
  1991, \aj, 102, 744

\bibitem[2003]{yi03}
  Yi, S. K., Kim, Y., \& Demarque, P.
  2003, \apjs, 144, 259 (Y$^2$)

\bibitem[2003]{yong03}
  Yong, D., \& Lambert, D. L.
  2003, \pasp, 115, 22

\end{thebibliography}
\end{document}